\documentclass[manuscript,screen]{acmart}

\setcopyright{acmlicensed}
\copyrightyear{2020}
\acmYear{2020}
\acmDOI{00.0000/0000000.0000000}

\acmJournal{TOPLAS}
\acmVolume{0}
\acmNumber{0}
\acmArticle{0}
\acmMonth{8}

\usepackage{hyperref}
\hypersetup{
    colorlinks=false,
}

\usepackage{listings}
\usepackage{float}
\newfloat{lstfloat}{htbp}{lop}
\floatname{lstfloat}{Listing}

\usepackage{caption}
\usepackage{subcaption}

\usepackage{manyfoot}

\usepackage{bm}

\newfloat{grammar}{h}{Grammar}

\usepackage{graphicx}

\usepackage{booktabs}

\begin{document}

\title[Pika Parsing]{Pika parsing: reformulating packrat parsing as a dynamic programming algorithm solves the left recursion and error recovery problems}

\author{Luke A. D. Hutchison}
\email{luke.hutch@alum.mit.edu}
\orcid{0000-0002-2937-8619}
\affiliation{\institution{Massachusetts Institute of Technology}}


\begin{abstract}
A recursive descent parser is built from a set of mutually-recursive functions, where each function directly implements one of the nonterminals of a grammar. A packrat parser uses memoization to reduce the time complexity for recursive descent parsing from exponential to linear in the length of the input. Recursive descent parsers are extremely simple to write, but suffer from two significant problems: (i) left-recursive grammars cause the parser to get stuck in infinite recursion, and (ii) it can be difficult or impossible to optimally recover the parse state and continue parsing after a syntax error. Both problems are solved by the \emph{pika parser}, a novel reformulation of packrat parsing as a dynamic programming algorithm, which requires parsing the input in reverse: bottom-up and right to left, rather than top-down and left to right. This reversed parsing order enables pika parsers to handle grammars that use either direct or indirect left recursion to achieve left associativity, simplifying grammar writing, and also enables optimal recovery from syntax errors, which is a crucial property for IDEs and compilers. Pika parsing maintains the linear-time performance characteristics of packrat parsing as a function of input length. The pika parser was benchmarked against the widely-used Parboiled2 and ANTLR4 parsing libraries, and the pika parser performed significantly better than the other parsers for an expression grammar, although for a complex grammar implementing the Java language specification, a large constant performance impact was incurred per input character for the pika parser, which allowed Parboiled2 and ANTLR4 to perform significantly better than the pika parser for this grammar (in spite of ANTLR4's parsing time scaling between quadratically and cubically in the length of the input with the Java grammar). Therefore, if performance is important, pika parsing is best applied to simple to moderate-sized grammars, or to very large inputs, if other parsing alternatives do not scale linearly in the length of the input. Several new insights into precedence, associativity, and left recursion are presented.
\end{abstract}

\ccsdesc[500]{Software and its engineering~Parsers}
\ccsdesc[400]{Theory of computation~Grammars and context-free languages}

\keywords{parsing, recursive descent parsing, top-down parsing, bottom-up parsing, PEG parsing, PEG grammars, packrat parsing, precedence, associativity, left-recursive grammars, parsing error recovery, memoization, dynamic programming}

\maketitle

\section{Introduction}

\subsection{Parsing generative grammars}

Since the earliest beginnings of computer science, researchers have sought to identify robust algorithms for parsing formal languages~\citep{aho1972,aho2006}. Careful work to establish the theory of computation provided further understanding into the expressive power of different classes of languages, the relationships between different classes of languages, and the relative computational power of algorithms that can parse different classes of languages~\citep{sipser2012}.

Generative systems of grammars, particularly Chomsky's context-free grammars (CFGs)~\citep{chomsky1956} and Kleene's regular expressions~\citep{kleene1951} became the workhorses of parsing for many data formats, and have been used nearly ubiquitously over several decades. However, inherent ambiguities and nondeterminism in allowable grammars increase the implementation complexity or decrease the runtime efficiency of parsers for these classes of language~\citep{lang1974,tomita2013}, due to parsing conflicts and/or exponential blowup in parsing time or space due to nondeterminism~\citep{kleene1951,mcnaughton1960}.

\subsection{Bottom-up parsers}

Much early work in language recognition focused on \emph{bottom-up parsing}, in which, as the input is processed in order, a forest of parse trees is built upwards from terminals towards the root. These parse trees are joined together into successively larger trees as higher-level \emph{productions} or \emph{reductions} are applied. This parsing pattern is termed \emph{shift-reduce}~\citep{aho2006}. Examples of shift-reduce parsers include early work on precedence parsers and operator precedence parsers~\citep{samelson1960,knuth1962,gray1969,aho1975,aho1977}, Floyd's operator precedence parsers (which are a proper subset of operator precedence parsers)~\citep{floyd1963,ruvzivcka1979}, the later development of LR~\citep{knuth1965} and LALR (lookahead)~\citep{deremer1969} parsers, the more complex GLR parser~\citep{lang1974} designed to handle ambiguous and nondeterministic grammars, and ``recursive ascent'' parsers~\citep{thomas1986}. Some parsers in this class run into issues with shift-reduce conflicts and/or reduce-reduce conflicts caused by ambiguous grammars; the GLR parser requires careful and complex data structure maintenance to avoid exponential blowup in memory usage, and to keep parsing time complexity to $O(n^3)$ in the length of the input. Optimal error recovery in shift-reduce parsers can be complex and difficult to implement.

\subsection{Top-down parsers}

More recently, \emph{recursive descent parsing} or \emph{top-down parsing} has increased in popularity, due to its simplicity of implementation, and due to the helpful property that the structure of mutually-recursive functions of the parser directly parallels the structure of the corresponding grammar.

The Pratt parser~\citep{pratt1973,crockford2007} is an improved recursive descent parser that combines the best properties of recursive descent and Floyd's Operator Precedence parsing, and has good performance characteristics. However the association of semantic actions with tokens rather than rules in a Pratt parser makes the parser more amenable to parsing of dynamic languages than static languages~\citep{crockford2007}.

Recursive descent parsers have suffered from three significant problems:

\begin{enumerate}
    \item {\bf Problem:} In the worst case, the time complexity of a na\"{\i}ve recursive descent parser scales exponentially in the length of the input, due to unlimited lookahead, and exponentially in the size of the grammar, due to the lack of termination of recursion at already-visited nodes during depth-first search over the Directed Acyclic Graph (DAG) of the grammar.\\
    {\bf Solution:} The discovery of \emph{memoized parsing} (formerly termed ``chart parsing''; now more often termed ``packrat parsing'' in the top-down case) reduced the worst case time complexity for recursive descent parsing from exponential to linear in the length of the input~\citep{norvig1991}. Packrat parsers employ memoization to avoid duplication of work, effectively breaking edges in the recursive descent call graph so that they traverse only a minimum spanning tree of the DAG-structured call graph of the parse, terminating recursion at the second and subsequent incoming edge to any recursion frame.
    \item {\bf Problem:} Without special handling~\citep{medeiros2014}, left-recursive grammars cause na\"{\i}ve recursive descent parsers to get stuck in infinite recursion, resulting in stack overflow.\\
    {\bf Prior progress:} It was found that direct or indirect left recursive rules could be rewritten into right recursive form (with only a weak equivalence to the left recursive form)~\citep{ford2002}, although the rewriting rules can be complex, especially when semantic actions are involved, and the resulting parse tree can diverge quite significantly from the structure of the original grammar, making it more difficult to create an abstract syntax tree from the parse tree. With some extensions, packrat parsing can be made to handle left recursion~\citep{ford2002,frost2006,frost2007,warth2008}, although usually with loss of linear-time performance, which is one of the primary reasons packrat parsing is chosen over other parsing algorithms~\citep{warth2008}. Some workarounds to supporting left recursion in recursive descent parsers only handle indirect left recursion, not direct left recursion.
    \item {\bf Problem:} If there is a syntax error partway through a parse, it is difficult or impossible to optimally recover the parsing state to allow the parser to continue parsing beyond the error. This makes recursive descent parsing particularly problematic for use in Integrated Development Environments (IDEs), where syntax highlighting and code assist need to continue to work after syntax errors are encountered during parsing. Recursive descent parsers are also not ideal for use in compilers, since compilers ideally need to be able to show more than one error message per compilation unit.\\
    {\bf Prior progress:} Many efforts have been made to improve the error recovery capabilities of recursive descent parsers, for example by identifying synchronization points from which parsing can recover with some reliability~\citep{medeiros2018}, or by employing heuristic algorithms to attempt to recover after an error, yielding an ``acceptable recovery rate'' from 41\% to 76\% in recent work~\citep{medeiros2020}. Up to and including this most recent state-of-art, optimal error recovery has resisted solution for recursive descent parsers.
\end{enumerate}

\subsection{\label{sec:RTL}Dynamic programming parsers (chart parsers)}

A chart parser is a type of parser suited to parsing ambiguous grammars~\citep{kay1980}. Chart parsers avoid exponential blowup in parsing time arising from the nondeterminism of a grammar by reducing duplication of work through the use of memoization. Top-down chart parsers (such as packrat parsers) use memoized recursion, whereas bottom-up chart parsers more specifically use dynamic programming (Section~\ref{sec:Levenshtein}). Chart parsers are comparable to the Viterbi algorithm~\citep{viterbi1967}, finding an optimal tree of paths through a directed acyclic graph (DAG) of all possible paths representing all possible parses of the input with the grammar. They may work unidirectionally or bidirectionally to build the parse tree.

The Earley parser is a top-down chart parser, and is mainly used for parsing natural language in computational linguistics~\citep{earley1970}. It can parse any context-free grammar, including left-recursive grammars. The Earley parser executes in cubic time in the general case, quadratic time for unambiguous grammars, and linear time for all LR(k) grammars -- however for most computer languages, Earley parsers are less efficient than alternatives, so they do not see much use in compilers. There is also some complexity involved in building a parse tree using an Earley parser, and Earley parsers may have problems with some nullable grammars. To increase efficiency, modern Earley parsing variants use a forward pass, to keep track of all active rules or symbols in the grammar at the current position, followed by a reverse pass, following all back-references in order to assemble the actual parse tree. Earley parsing algorithms are not directly extensible to handle PEG (Section~\ref{sec:peg_grammars}), as a result of PEG supporting arbitrary lookahead. The Earley parser may be converted from top-down memoized recursive form into bottom-up dynamic programming form~\citep{voisin1988}.

``Parsing with pictures'' is a chart parsing algorithm that provides an alternative approach to parsing context-free languages. The authors claim that this method is simpler and easier to understand than standard parsers using derivations or pushdown automata~\citep{pingali2012}. This parsing method unifies Earley, SLL, LL, SLR, and LR parsers, and demonstrates that Earley parsing is the most fundamental Chomskyan context-free parsing algorithm, from which all others derive.

A particularly interesting though inefficient bottom-up chart parser is the CYK algorithm for context-free grammars~\citep{cocke1970,younger1967,kasami1966}. This parsing algorithm employs dynamic programming to efficiently recursively subdivide the input string into all substrings, and determines whether each substring can be generated from each grammar rule. This results in $O(n^3 |G|)$ performance in the length of the input $n$ and the size of the grammar $|G|$. The algorithm is robust and simple, and has optimal error recovery capabilities since all rules are matched against all substrings of the input. However, the CYK algorithm is inefficient for very large inputs due to scaling as the cube of the input length $n$, and additionally, the algorithm requires the grammar to be reduced to Chomsky Normal Form (CNF)~\citep{chomsky1959} as a preprocessing step, which can result in a large or even an exponential blowup in grammar size in the worst case -- therefore in practice, $|G|$ may also be large.

\subsection{Left-to-right vs. right-to-left parsing}

In prior work, with the exception of left/right orderless parsers such as the CYK algorithm, and parsers that perform a right-to-left ``fixup pass'' as a postprocessing step such as modern Earley parsers, almost all parsers -- and certainly all practical parsers -- have consumed input in left-to-right order only.

In fact, reversing the direction in which input is consumed from left-to-right to right-to-left changes the recognized language for most parsers applied to most grammars. This is principally because most parsing algorithms are only defined to operate either top-down or bottom-up, and even if the left-to-right parsing order is reversed, the top-down vs. bottom-up order cannot be easily reversed for most parsing algorithms, because this order is fundamental to the design of the algorithm. Without reversing \emph{both} parsing axes, the semantics of a ``half-reversed'' parsing algorithm will not match the original algorithm (Section~\ref{sec:DP_order}).

Ford briefly proposed a bottom-up, right-to-left parsing algorithm in the \emph{Tabular Top-Down Parsing} section of~\citep{ford2002} and~\citep{ford2002b}, yet quickly concluded that \emph{``An obvious practical problem with the tabular right-to-left parsing algorithm above is that it computes many results that are never needed. An additional inconvenience is that we must carefully determine the order in which the results for a particular column are computed, so that parsing functions ... that depend on other results from the same column will work correctly. Packrat parsing is essentially a lazy version of the tabular algorithm that solves both of these problems.''} The author then spent the bulk of the remaining manuscript describing top-down, left-to-right packrat parsing, and does not appear to have further pursued the bottom-up, right-to-left parsing approach. The two problems raised by Ford are addressed to a reasonable degree in pika parsing by minimizing the memoization of clauses that cannot possibly be part of the parse tree at a given position (Sections~\ref{sec:parentClauses},~\ref{sec:zerolen}), and through the use of topological sorting of clauses (Section~\ref{sec:topoOrder}).

\subsection{\label{sec:peg_grammars}Parsing Expression Grammars (PEG)}

A significant refinement of recursive descent parsing known as Parsing Expression Grammars (PEGs, or more commonly, but redundantly, ``PEG grammars'') were proposed by Ford in his 2002 Master's thesis~\citep{ford2002,ford2002b}. In contrast to most other parsers, which represent grammar rules as generative productions or derivations, PEG grammar rules represent \emph{greedily-recognized patterns,} replacing ambiguity with deterministic choice (i.e. all PEG grammars are unambiguous). There are only a handful of PEG rule types, and they can be implemented in a straightforward way. As a subset of all possible recursive rule types, all non-left-recursive PEG grammars can be parsed directly by a recursive descent parser, and can be parsed efficiently with a packrat parser. PEG grammars are more powerful than regular expressions, but it is conjectured (though as yet unproved) that there exist context-free languages that cannot be recognized by a PEG parser~\citep{ford2004}). Top-down PEG parsers suffer from the same limitations as other packrat parsers: they cannot parse left-recursive grammars without special handling, and they have poor error recovery properties.

A PEG grammar consists of a finite set of \emph{rules}. Each rule has the form $A \leftarrow e$, where $A$ is a unique \emph{rule name}, and $e$ is a \emph{parsing expression}, also referred to as a \emph{clause}. One rule is designated as the \emph{starting rule}, which becomes the root for top-down recursive parsing. PEG clauses are constructed from \emph{subclauses}, \emph{terminals}, and \emph{rule references}. The subclauses within a rule's clause may be composed into a tree, with \emph{PEG operators} forming the non-leaf nodes, and terminals or rule references forming the leaf nodes.

Terminals can match individual characters, strings, or (in a more complex parser) even regular expressions. One special type of terminal is the \emph{empty string} $\epsilon$ (with ASCII notation \texttt{()}), which always matches at any position, even beyond the end of the input string, and consumes zero characters.

The PEG operators (Table~\ref{tab:peg_operators}) are defined as follows:

\begin{itemize}
    \item A \texttt{Seq} clause (termed by Ford \emph{Sequence}~\citep{ford2002}) matches the input at a given start position if all of its subclauses match the input in order, with the first subclause match starting at the initial start position, and each subsequent subclause match starting immediately after the previous subclause match. Matching stops if a single subclause fails to match the input at its start position.
    \item A \texttt{First} clause (termed by Ford \emph{Ordered Choice}) matches if any of its subclauses match, iterating through subclauses in left to right order, with all match attempts starting at the current parsing position. After the first match is found, no other subclauses are matched, and the \texttt{First} clause matches. If no subclauses match, the \texttt{First} clause does not match. The \texttt{First} operator gives a top-down PEG parser \emph{limited backtracking capability}.
    \item A \texttt{OneOrMore} clause (termed by Ford \emph{Greedy Positive Repetition}) matches if its subclause matches at least once, greedily consuming as many matches as possible until the subclause no longer matches.
    \item A \texttt{NotFollowedBy} clause matches if its subclause does not match at the current parsing position, but it does not consume any characters if the subclause does not match. This operator provides \emph{lookahead, logically negated}.
\end{itemize}

Two additional PEG operators may be offered for convenience, and can be defined in terms of the basic PEG operators:

\begin{itemize}
    \item A \texttt{FollowedBy} clause matches if its subclause matches at the current parsing position, but no characters are consumed even if the subclause does match. This operator provides \emph{lookahead}. A \texttt{FollowedBy} clause may be transformed into a doubly-nested \texttt{NotFollowedBy} clause.
    \item An \texttt{Optional} clause matches if its subclause matches. If the subclause does not match, the \texttt{Optional} clause still matches, consuming zero characters (i.e. this operator always matches). An \texttt{Optional} clause $(e?)$ can be automatically transformed into \texttt{First} form $(e / \epsilon)$, reducing the number of PEG operators that need to be implemented directly.
    \item A \texttt{ZeroOrMore} clause (termed by Ford \emph{Greedy Repetition}) matches if its subclause matches zero or more times, greedily consuming as many matches as possible. If its subclause does not match (in other words if the subclause matches zero times), then the \texttt{ZeroOrMore} still matches the input, consuming zero characters (i.e. this operator always matches). A \texttt{ZeroOrMore} clause $(e*)$ can be automatically transformed into \texttt{Optional} form using a \texttt{OneOrMore} operator $((e+)?)$, which can be further transformed into \texttt{First} form using a \texttt{OneOrMore} operator $({e+} / \epsilon)$, again reducing the number of operators that need to be implemented.
\end{itemize}

\begin{table*}[tbp]
    \centering
    {\small
    \begin{tabular}{lcccl}\toprule
    \textbf{Name}          & \textbf{Num subclauses} & \textbf{Notation}     & \textbf{Equivalent to}                  & \textbf{Example rule in ASCII notation} \\\midrule
    \texttt{Seq}           & 2+                      & $e_1$ $e_2$ $e_3$     &                                         & \texttt{Sum <- Prod {\textquotesingle}+{\textquotesingle} Prod;} \\
    \texttt{First}         & 2+                      & $e_1$ / $e_2$ / $e_3$ &                                         & \texttt{ArithOp <- {\textquotesingle}+{\textquotesingle} / {\textquotesingle}-{\textquotesingle};} \\
    \texttt{OneOrMore}     & 1                       & $e+$                  &                                         & \texttt{Whitespace <- [ {\char`\\}t{\char`\\}n{\char`\\}r]+;} \\
    \texttt{NotFollowedBy} & 1                       & $!e$                  &                                         & \texttt{VarName <- !ReservedWord [a-z]+;} \\\midrule
    \texttt{FollowedBy}    & 1                       & $\&e$                 & $!!e$                                   & \texttt{Exclamation <- Word \&{\textquotesingle}!{\textquotesingle};} \\
    \texttt{Optional}      & 1                       & $e?$                  & $e / \epsilon$                          & \texttt{Value <- Name ({\textquotesingle}[{\textquotesingle} Num {\textquotesingle}]{\textquotesingle})?;} \\
    \texttt{ZeroOrMore}    & 1                       & $e*$                  & $(e+)?$ \textbf{[or]} ${e+} / \epsilon$ & \texttt{Program <- Statement*;} \\\bottomrule
    \end{tabular}
    }
    \caption{\label{tab:peg_operators}PEG operators, defined in terms of subclauses $e$ and $e_i$, and the empty string $\epsilon$.}
    \Description[PEG operators]{PEG operators, defined in terms of subclauses $e$ and $e_i$, and the empty string $\epsilon$.}
\end{table*}

\subsection{\label{sec:Levenshtein}Dynamic Programming: the Levenshtein distance} Dynamic programming (DP) is a method used to solve recurrence relations efficiently when direct recursive evaluation of a recurrence expression would recompute the same terms many times (often an exponential number of times). DP is analogous to an inductive proof, working bottom-up from base cases, with the addition of \emph{memoization}, or the storing of recurrence parameters and results in a table so that no specific invocation of a recurrence needs to be recursively computed more than once. One of the most commonly taught DP algorithms is the Levenshtein distance or string edit distance algorithm~\citep{levenshtein1966,kruskal1983}. This algorithm measures the number of edit operations (deletions, insertions, and modifications of one character for another) that are required to convert one string $A$ of length $N_A$ into another string $B$ of length $N_B$.

\begin{figure}[ht]
    \centering
    \includegraphics[width=1\textwidth]{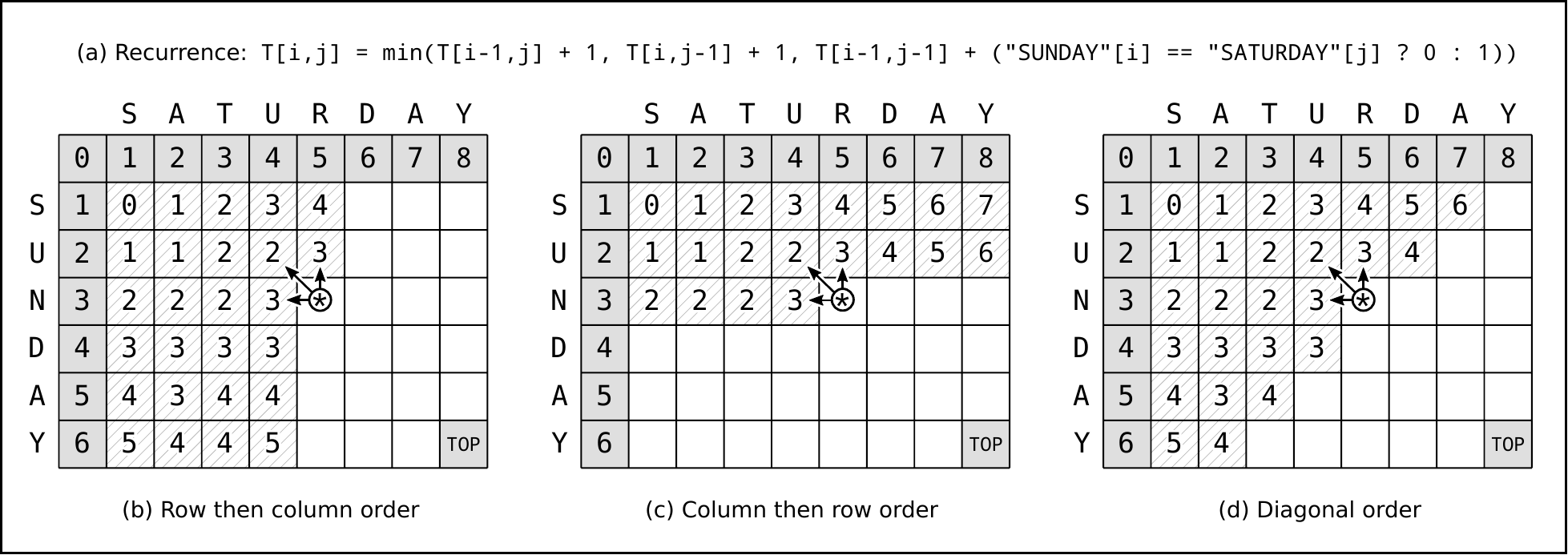}
    \caption{\label{fig:Levenshtein}Example of a canonical dynamic programming algorithm, the Levenshtein distance or string edit distance}
    \Description[Levenshtein distance]{Example of a canonical dynamic programming algorithm, the Levenshtein distance or string edit distance}
\end{figure}

The Levenshtein distance algorithm is depicted in Fig.~\ref{fig:Levenshtein}. The \emph{DP table} or \emph{memo table} for the algorithm consists of $(N_A+1)$ rows and $(N_B+1)$ columns, initialized by filling the first row and the first column with ascending integers starting with $0$ (these are the \emph{base cases} for induction or for the recurrence, marked with a gray background). The remaining entries in the table are populated top to bottom and left to right, until the toplevel recurrence evaluation is reached (marked ``\texttt{TOP}'' in the bottom right corner). The current cell value (marked with a circled asterisk) is calculated by applying the Levenshtein distance recurrence (Fig.~\ref{fig:Levenshtein}a), which is a function of the values in three earlier-computed table cells (shown with arrows): the cell in the same row but previous column, the cell in the the same column but previous row, and the cell in the previous column and previous row. After the value of the current cell is calculated, the value is stored in the DP table or \emph{memoized} for future reference.

The DP table must be populated in a reasonable order, such that a recurrence value is always stored in a table cell before any other cell that depends upon it is evaluated. For the Levenshtein recurrence, if an attempt were made to populate the DP table from right to left and/or bottom to top, then the recurrence would depend upon values that are not yet calculated or stored in the table, since all dependencies are to the left and/or above the current cell, therefore the algorithm would fail. In fact, all DP algorithms fail if the DP table is not populated in an order that respects the data dependency structure.

All Levenshtein recurrence dependencies are \emph{static} or fixed: the recurrence always refers to three previous table entries using the same exact dependency structure. This allows a correct population order to be determined simply and reliably by examination of the dependency structure. For this algorithm, many population orders are possible, including: row first then column (Fig.~\ref{fig:Levenshtein}b), column first then row (Fig.~\ref{fig:Levenshtein}c), and sweeping a ``wavefront'' from top left to bottom right, aligned with the trailing diagonal, over which cell values can be populated in any order, or even in parallel (Fig.~\ref{fig:Levenshtein}d). In fact many wavefront angles would work, as long as the wavefront moves from top left to bottom right (e.g. a wavefront with a gradient of 0.5 or 2 relative to the trailing diagonal will also populate the table in the correct order).

Note that as described, the Levenshtein distance algorithm works bottom-up, in the sense that ``bottom'' refers to the base cases of recursion (i.e. the top row and the leftmost column of the table) and ``up'' refers to moving towards the toplevel recurrence evaluation (marked ``\texttt{TOP}'' in the bottom-right corner of the table): moving from top left to bottom right in the DP table corresponds to moving upwards in the recursion hierarchy. However, this same recurrence could have been solved just as easily using memoized recursive descent instead: recursing down the tree of recurrence frames from the topmost recurrence evaluation to the base cases, checking whether an entry in the memo table has already been computed and memoized before recursing to it, and memoizing values as they are computed. The computed result of this top-down memoized variant of the Levenshtein distance algorithm would be the same as the bottom-up dynamic programming algorithm. In both variants, the time complexity would be $O(N_A N_B)$.

Similarly, memoized recursive descent parsing, or packrat parsing, can be inverted to form a bottom-up dynamic programming equivalent, newly introduced here as pika parsing.

\section{\label{sec:thePikaParser}Pika parsing: A DP algorithm for bottom-up, right to left parsing}

\subsection{Overview of pika parsing}

The \emph{pika parser}\footnote{A pika (``PIE-kah'') is a small, mountain-dwelling, rabbit-like mammal found in Asia and North America, typically inhabiting moraine fields near the treeline. Like a packrat, it stores food in caches or ``haystacks'' during the warmer months to survive the winter.} is a new memoizing parsing algorithm that uses dynamic programming (DP) to apply PEG operators in reverse order relative to packrat parsing, i.e. bottom-up (from terminals towards the root of the parse tree) and from right to left (from the end of the input back towards the beginning). This is the reverse of the top-down, left-to-right order of standard packrat parsing, in the same sense that the Levenshtein distance recurrence can be optimally computed as either the original bottom-up DP algorithm or as a top-down memoized recursive descent algorithm (Section~\ref{sec:Levenshtein}).

As with packrat parsing, pika parsing does not need a lexer or lex preprocessing step to parse an input: the parser can perform both the lexing and parsing steps of a traditional parsing algorithm.

Pika parsing scales linearly in the length of the input, which is rare among parsing algorithms, and has favorable performance characteristics for smaller grammars and long inputs, though may incur a large constant cost per input character for large and complex grammars (Section~\ref{sec:Performance}). In particular, the pika parsing algorithm has essentially the same parsing and error recovery capabilities as the CYK parsing algorithm, but with cubically better scaling characteristics.

Pika parsers and packrat parsers are not semantically equivalent, since pika parsers support left-recursive grammars directly (Section~\ref{sec:LeftRecursion}), whereas standard packrat parsers do not. In this sense, pika parsers are more flexible than standard packrat parsers, because pika parsers can recognize grammars that standard packrat parsers cannot. Note however that since every left-recursive grammar can be transformed into non-left-recursive form~\citep{ford2002}, the language recognized by pika parsers and packrat parsers is the same -- although the resulting parse tree may deviate far from the syntactic structure of the document in the results of packrat parsing, if left-recursive rules have to be rewritten into non-left-recursive form. Pika parsers also support optimal error recovery (Section~\ref{sec:ErrorRecovery}), which so far has not proved tractable for packrat parsers.

\subsection{The choice of the PEG formalism}

The pika parser is defined in this paper to work with PEG grammars (Section~\ref{sec:peg_grammars}), which are unambiguous and deterministic, eliminating the ambiguity and nondeterminism that complicate other bottom-up parsers and increase parsing time complexity. However, other unambiguous and deterministic grammar types could also be used, as long as they depend only upon matches to the right of or below a given entry in the memo table (Section~\ref{sec:DP_order}).

\subsection{PEG grammars are recurrence relations, so can be solved with DP}

Recursive descent PEG parsing hierarchically applies matching rules, which may reference other rules in a graph structure (possibly containing cycles). Recursive descent parsing works exactly like the hierarchical evaluation of a recurrence relation, however for recursive descent parsing, the dependencies between recurrence frames are \emph{dynamic:} any \texttt{Seq} or \texttt{OneOrMore} clause is able to match subclause(s) either at a given start position, or any distance to the right of the start position, depending on how many characters are consumed by each successive subclause match.

\begin{figure}[ht]
    \centering
    \includegraphics[width=1\textwidth]{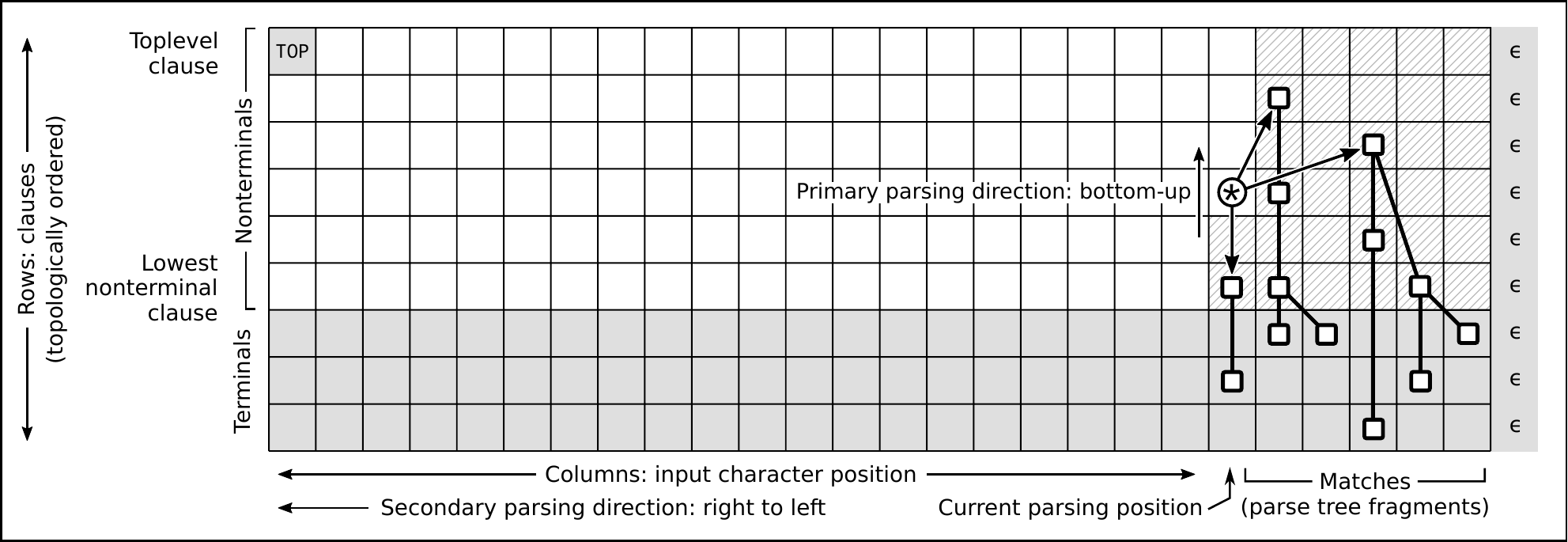}
    \caption{\label{fig:ParsingRecurrence}Packrat parsing reformulated as a dynamic programming algorithm}
    \Description[Parsing as dynamic programming]{Recursive descent parsing, reformulated into dynamic programming form}
\end{figure}

In spite of the need for dynamic dependencies when using recursive descent parsing with PEG grammars, it is still possible to formulate PEG parsing as a DP algorithm (Fig.~\ref{fig:ParsingRecurrence}). A DP parser creates a memo table with a row for each clause or subclause in the grammar, sorted topologically so that terminals are represented by rows at the bottom of the table (these are base cases for the recurrence), and the toplevel clause is represented by the first row (Section~\ref{sec:topoOrder}). The memo table has one column per character in the input, and the empty string $\epsilon$ can be considered to match beyond the end of the string (which creates a column of base cases for the recurrence to the right of the table). The top-leftmost cell (marked \texttt{TOP} in Fig.~\ref{fig:ParsingRecurrence}) is the toplevel recurrence frame, corresponding to the entry point for recursive descent parsing. As the table is populated, parse tree fragments are grown or extended upwards from leaves consisting of terminal matches. Parse tree fragments are extended upwards when a clause in the grammar is successfully matched at a given input position, meaning that a match was found in the memo table for each of the requisite subclauses of the clause. Thus each matching subclause represents a node in the generated parse tree (indicated with small white boxes in Fig.~\ref{fig:ParsingRecurrence}). The current memo entry is marked with a circled asterisk in Fig.~\ref{fig:ParsingRecurrence}, and memo entries that are fully parsed and finalized (or that cannot represent matches of grammar rules against the input) are shown with a hatched background.

\subsection{\label{sec:DP_order}Correct memo table population order for a DP PEG parser: bottom-up, right to left}

When populating any entry in the memo table, the clause given by the row can depend upon subclause matches represented by memo entries either \emph{below} the current cell (corresponding to a top-down recursive call from a higher-level/lower-precedence clause to a lower-level/higher-precedence clause in the grammar), or upon cells \emph{in any row of any column to the right of the current parsing position} (corresponding to the second or subsequent subclause matches of a \texttt{Seq} or \texttt{OneOrMore} clause, after the first subclause match consumed at least one character).

By examining this dependency relationship, it is immediately clear that there is only one correct order in which to populate the memo table so that dependencies are always memoized before they are needed: bottom-up, then right to left. In other words, any parse nodes in the current column should be extended upwards as far as possible before the parser moves to the previous column. Without populating the memo table in this order, parsing will fail to complete, because the parser will attempt to read memo table entries for matches that have not yet been performed.

Bottom-up parsing is not unusual, but parsing from right to left is highly unusual among parsing algorithms (Section~\ref{sec:RTL}). Nevertheless, this is the only DP table population order that can successfully convert a memoized recursive descent parser into a DP or chart parser.

Note that when parsing from right to left, rules are still applied left-to-right. Therefore, there is no wasted work triggered by a spurious match of the right-most subclause of a \texttt{Seq} clause, for example by matching the closing parenthesis in a clause like \texttt{(\textquotesingle{}(\textquotesingle{} E \textquotesingle{})\textquotesingle{})}: a \texttt{Seq} clause is not matched until its \emph{leftmost} subclause is found to match, at which point the second and subsequent subclause matches are simply looked up in the memo table.

\subsection{\label{sec:topoOrder}Topological sorting of clauses into bottom-up order}

In order to be able to invert the top-down call order of a recursive descent parser so that rules can be parsed bottom-up, the grammar graph must be sorted into bottom-up topological order, from leaves (terminals) to root (the toplevel rule).

Topological order is straightforward to obtain for an arbitrary tree or DAG: simply perform a depth-first search (DFS) of the graph, writing the node to the end of the bottom-up topological order (or to the beginning of the top-down topological order) right before exiting a recursion frame (i.e. the topological sort order is the deduplicated postorder traversal order of the nodes in a DAG). If there is more than one greatest upper bound or toplevel node in the DAG, the DFS used to produce the topological sort order needs to be applied to each toplevel node in turn, appending to the end of the same topological sort output for each DFS call, and reusing the same set of visited nodes for each call (so that descendant nodes are not visited twice across different DFS calls).

However, the topological sort order is not defined for a graph containing cycles. To produce a reasonable topological sort order for a grammar graph containing cycles, the cycles must be broken in appropriate places to form a tree or DAG. The appropriate way to break cycles for a grammar graph is to remove the edge that causes a path from a toplevel node to double back to an node already in the path, since the node that would otherwise appear twice in the path must be the entry point for the cycle (and/or it must be the lowest precedence clause in the cycle).

The topological order for a set of grammar rules is obtained by running DFS starting from each node in the following order, building a single topological order across all DFS invocations, and sharing the set of visited nodes across invocations:

\begin{enumerate}
    \item All toplevel nodes, i.e. any rules that are not referred to by any other rule. (There will typically be only one of these in a grammar.)
    \item The lowest precedence clause in each precedence hierarchy. (The lowest precedence clause is the entry point into a precedence hierarchy.)
    \item ``Head nodes'' of any other cycles in the grammar: this is the set of all nodes that can be reached twice in any path that can be traversed through the grammar starting from a toplevel node (halting further recursion when a head node is reached). A cycle detection variant of DFS is used to find these head nodes.
\end{enumerate}

This topological sort algorithm labels each clause with an index (stored in the \texttt{clauseIdx} field of the \texttt{Clause} class -- Section~\ref{sec:Clause}) representing the position of the clause in the topological order of all clauses in the grammar, in increasing order from terminals up to the toplevel clause.

The implementation of this topological sort algorithm can be seen in Listing~\ref{lst:topoSort}.

\subsection{\label{sec:parentClauses}Scheduling parent clauses for evaluation}

To extend the memoized parse tree fragment upwards from the current memo entry, each time a clause \texttt{X} successfully matches the input, the parser must identify all parent clauses for which \texttt{X} is the first subclause -- or, in the case of a \texttt{Seq} operator, for which \texttt{X} is not the first subclause, but \texttt{X} would be matched in the same start position as the parent clause (as a result of all previous sibling subclauses matching but consuming zero characters). These parent clauses are termed the \emph{seed parent clauses} of the subclause, and since the grammar does not change during parsing, seed parent clauses can be determined for each clause during initialization. Seed parent clauses for \texttt{X} would include \texttt{(W / X / Y)}, \texttt{(X*)}, \texttt{(X?)}, \texttt{(X Y Z)} and \texttt{(W? X Z)}, but would not include \texttt{(W X Z)} if a match of \texttt{W} always consumes at least one character, since \texttt{X} would then match at a start position strictly to the right of the start position of a match of its parent clause. As the parent clause \texttt{(W X Z)} will always be triggered by a match of \texttt{W} in the same start position before the memo table is even checked for a match of \texttt{X}, \texttt{X} does not need to trigger the parent clause.

Whenever a new match is found for the clause corresponding to the row of the current memo entry, all seed parent clauses of the matching clause are scheduled for evaluation at the same start position (i.e. in the same column as the current entry). This expands the ``DP wavefront'' upwards.

Even though all memo table entries to the right of and below the current cell are finalized (shown as hatched in Fig.~\ref{fig:ParsingRecurrence}), not every memo table entry needs to be evaluated, since no typical grammar will give rise to a parse tree consisting of one node for every entry in the memo table -- therefore, the memo table can and should be stored sparsely. The seed parent clauses mechanism improves memory consumption and performance of the pika parser, since it allows a pika parsing implementation to use a sparse map for the memo table, and evaluate only the clauses that can possibly match at a given position.

\subsection{\label{sec:handlingCycles}Adjustment of memo table population order to handle cycles}

When traversing down any path in a recursive descent parse tree from the root, precedence increases downwards towards the terminals: the start rule or toplevel node at the top of the parse tree has the lowest precedence, and leaves or terminal matches at the bottom of the parse tree have the highest precedence. It is therefore common to write recursive descent grammars in \emph{precedence climbing} form (Section~\ref{sec:precedence}), where if a rule fails to match, it fails over to a higher level of precedence. Since pika parsers work bottom-up, they are actually \emph{precedence descent} parsers, even though the grammar rules are still defined in precedence climbing form, i.e. as if designed to be parsed by a top-down parser. (Actually, although the DP table population order is bottom-up for pika parsing, the memo table lookup step where each clause is evaluated to see if it matches technically applies the match top-down, just as the Levenshtein distance DP algorithm is bottom-up, but applies the recurrence top-down in each DP table lookup step.)

When a memo entry depends upon a another entry to the right of the current column and \emph{above} the current row (as shown for two of the three dependency arrows in Fig.~\ref{fig:ParsingRecurrence}), this indicates that a \emph{decrease} in precedence, which typically only occurs through the use of a \emph{precedence override pattern,} such as a rule that matches the lowest precedence clause of a precedence hierarchy surrounded by parentheses. Any dependency upon a topologically-higher clause also indicates that a grammar cycle has been encountered, since the clauses in the rows of the memo table were sorted topologically, breaking cycles only at the point where a path doubles back to a lower-precedence clause (Section~\ref{sec:topoOrder}).

When a grammar contains cycles, to replicate the left-recursive semantics of any top-down parser that can handle left recursion, a bottom-up parser must maximally expand the parse tree upwards for nodes that form part of a cycle before moving further upwards to populate the memo table entries higher in the same memo table column. This is accomplished by using a priority queue to schedule parent clauses for matching. The priority queue sorts in increasing order of bottom-up topological sort index of the clause (with the lowest topological sort indexes being assigned to the terminals, and the highest topological sort index being assigned to the start rule or toplevel rule). If a parent clause in a grammar cycle has lower precedence, then another match of clauses within the cycle will be attempted before moving further up the column, as a direct consequence of the use of a priority queue to schedule clauses for matching. This behavior enables left recursion (and right recursion) in grammar clauses to be handled directly.

\subsection{\label{sec:leftRecursionTermination}Ensuring termination when grammars contain cycles}

Recursive descent parsers get stuck in infinite recursion if grammar rules are left-recursive. To ensure termination of a pika parser, whenever a new match is found for a memo entry that already contains a match (which only happens if a cycle has been traversed in the grammar), the newer match must be longer than the older match for the memo entry to be updated, and for seed parent clauses to be scheduled for evaluation.

In other words, looping around a cycle in the grammar must consume at least one additional character per loop. This has two effects:

\begin{itemize}
    \item \emph{Left recursion is guaranteed to terminate at some point}, because the input string is of finite length.
    \item \emph{Left recursion is able to terminate early (before consuming the entire input)} whenever looping around a cycle in the grammar results in a match of the same length as the previous match, or a mismatch, indicating that no higher-level matches were able to be found.
\end{itemize}

There is one exception to this: for \texttt{First} clauses, even before the length of the new match is checked against the length of the existing match as described above, the index of the matching subclause of each match must be compared, since the semantics of the \texttt{First} PEG operator require that an earlier matching subclause take priority over a later matching subclause (Section~\ref{sec:Match}).

If an entry in the memo table already contains a match, and a new and better match is found for the same memo entry due to a cycle in the grammar, then the older match is linked into the parse tree as a descendant node of the new match, so the older match is not lost due to being overwritten by the newer match in the memo entry.

\subsection{Step-through of pika parsing}

\begin{figure}[ht]
    \centering
    \includegraphics[width=1\textwidth]{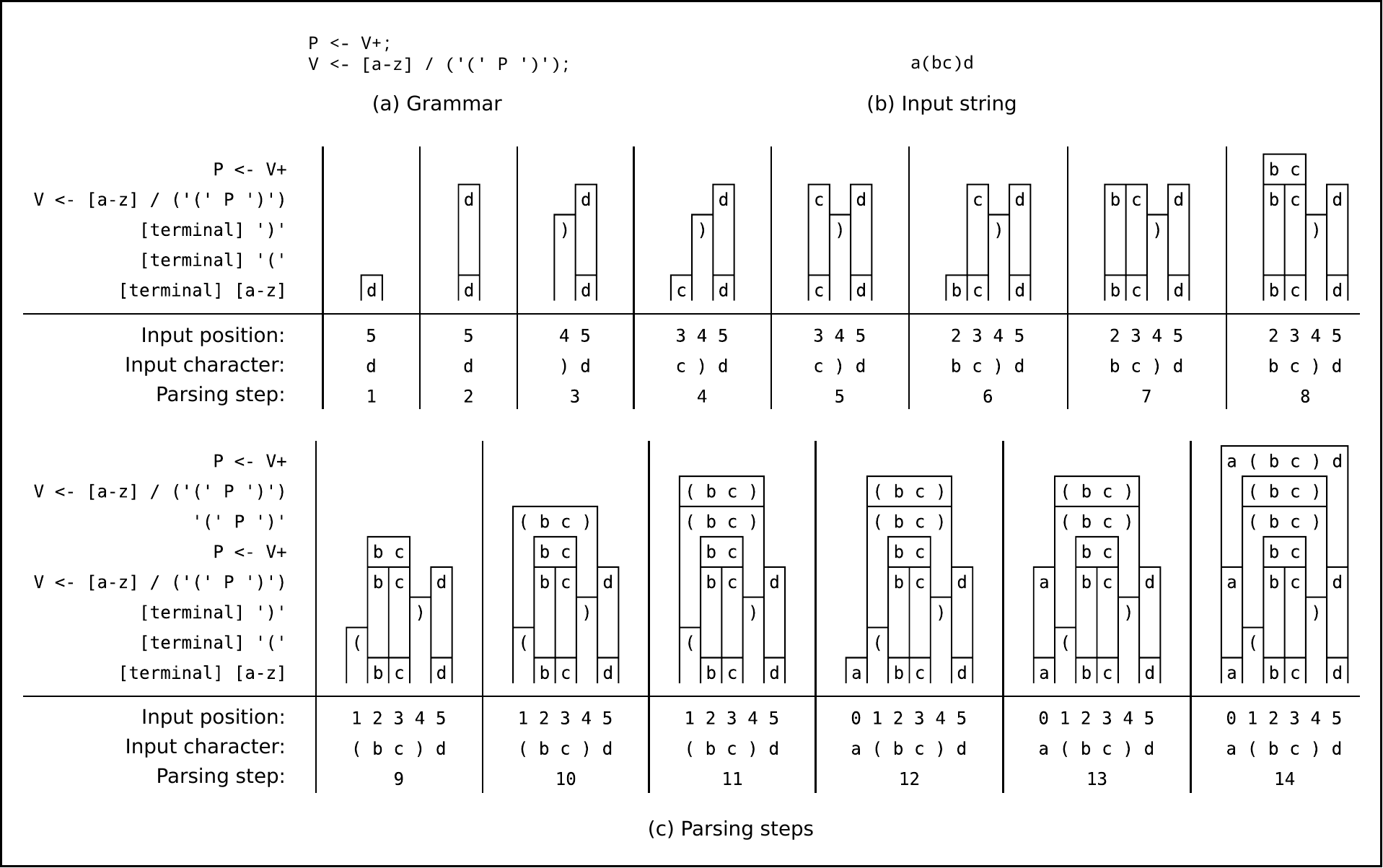}
    \caption{\label{fig:ParsingSteps}Example of pika parsing, with each successive state of the memo table shown as a separate step}
    \Description[Parsing steps]{Example of pika parsing, with each successive state of the memo table shown as a separate step}
\end{figure}

Fig.~\ref{fig:ParsingSteps} illustrates how pika parsing works for a concrete grammar and input, broken down into the 14 separate steps needed to parse the input. In step 1, the parsing position is the last character position, i.e. 5, and the terminal \texttt{[a-z]} matches in this position, consuming one character. There is only one seed parent clause of the terminal clause \texttt{[a-z]}, which is the rule \texttt{V}, so \texttt{V} is scheduled to be matched at position 5 in in step 2. In step 2, rule \texttt{V} matches its first subclause \texttt{[a-z]}, so the whole rule \texttt{V} is found to match at position 5, also consuming one character.

There is one seed parent clause of \texttt{V}, which is \texttt{P}, and since \texttt{V} matches at position 5, \texttt{(P <- V+)} is scheduled to be matched at position 5, and is also found to match at position 5, consuming one character. However, this match of \texttt{P} is not depicted in this diagram, because the match is not picked up as a subclause match by higher clauses, so it represents a spurious match, and is not linked into the final parse tree when parsing is complete.

Once no more upwards expansion is possible in input position 5, the parser moves to the left, to input position 4, and again tries matching all terminals in the bottom rows of the table, in order to trigger the bottom-up growth of the parse tree, then moves to successively higher clauses by following their seed parent clauses.

This process continues, until in step 10 and 11 the upwards growth in the parse tree results in a \emph{loop} around the single cycle in the grammar rules: rule \texttt{V} matches the string \texttt{"(bc)"}, but \texttt{V} also matches each of \texttt{"b"} and \texttt{"c"}. There is only one row in the memo table for each grammar clause, so to allow the whole structure of the tree to be visualized simply, if a parse tree fragment loops around a cycle in the grammar, then clauses involved in the loop are shown duplicated on different rows.

A second loop around the grammar cycle can be observed in the final step: rule \texttt{P} matches the entire input, but also matches \texttt{"bc"}.

\section{Properties of pika parsers}

\subsection{\label{sec:LeftRecursion}Support for left and right recursion}

With the conditions described in Section~\ref{sec:leftRecursionTermination} in place, it is possible to handle both left and right recursion, either direct or indirect, using a single simple parsing mechanism. (It should be pointed out that reversing the parsing direction from left-to-right to right-to-left does not simply replace the left recursion problem with a corresponding right recursion problem.) No additional special handling is required to handle grammar cycles of any degree of nested complexity.

\begin{figure}[ht]
    \centering
    \includegraphics[width=1\textwidth]{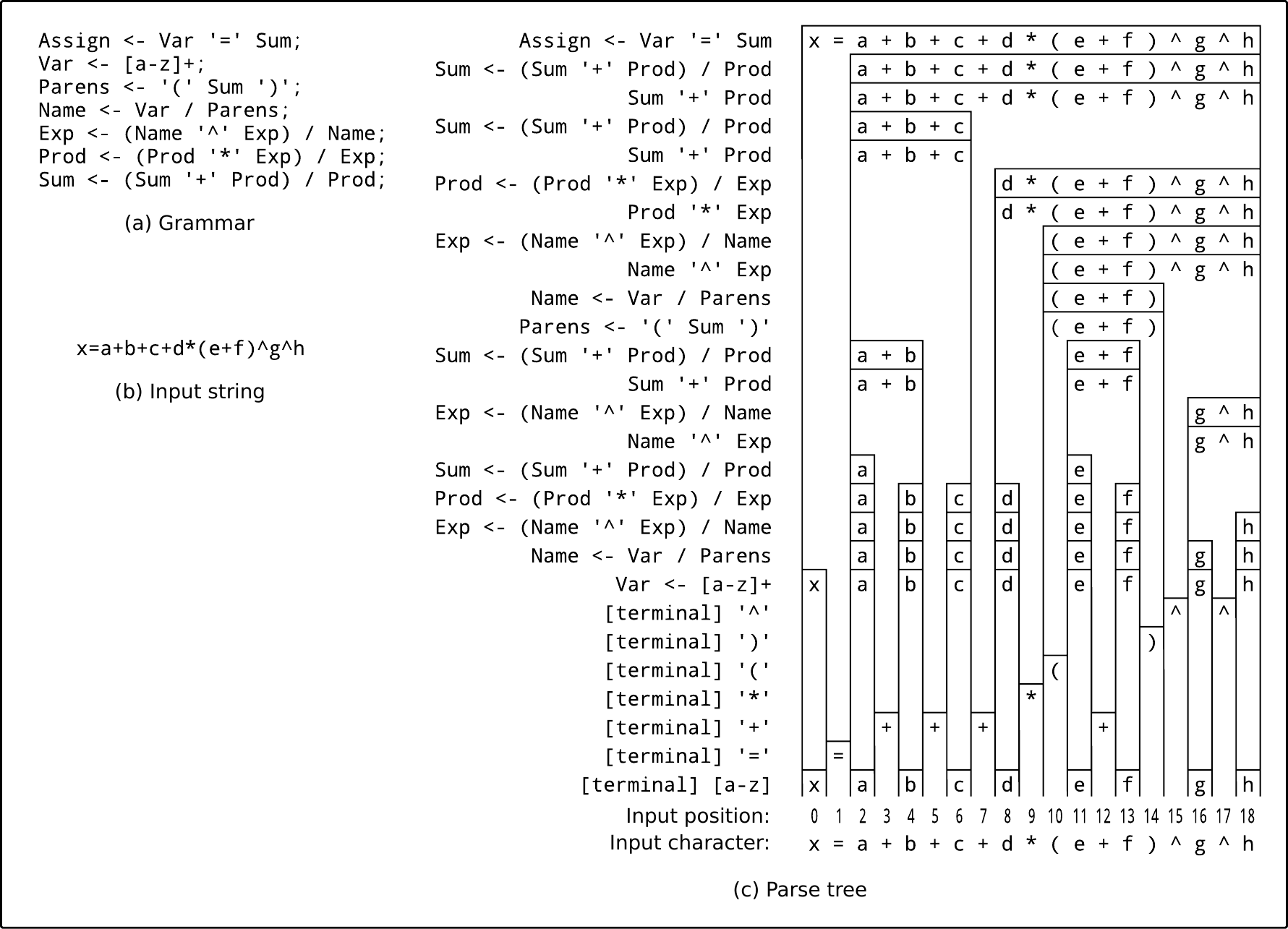}
    \caption{\label{fig:LeftAndRightRecursiveRule}Example of parsing of left- and right-recursive rules within the same multi-precedence grammar}
    \Description[Parsing left- and right-recursive rules]{Example of parsing of left- and right-recursive rules within the same multi-precedence grammar}
\end{figure}

An example of applying pika parsing to a grammar that contains both left-recursive rules (addition or \texttt{Sum}, and multiplication or \texttt{Prod}) and a right-recursive rule (exponentiation or \texttt{Exp}) is given in Fig.~\ref{fig:LeftAndRightRecursiveRule}. Note that all precedence and associativity rules are correctly respected in the final parse tree: the run of equal-precedence addition terms is built into a left-associative parse subtree; the run of equal-precedence exponentiations is built into a right-associative parse subtree; exponentiation takes precedence over multiplication, which takes precedence over addition (except in the case of the use of parentheses to override the precedence order for the inner summation, \texttt{(e+f)}). The parse terminates after several different loops around cycles in the grammar reach the toplevel rule, \texttt{Assign}, consuming the entire input.

\subsection{\label{sec:ErrorRecovery}Error recovery in pika parsers}

PEG rules only depend upon the span of input from their start position to their end position, in other words they do not depend upon any characters before their start position or after their end position. Because the input is parsed from right to left, meaning the memo table is always fully populated to the right of the current parse position, the span of input to the right of a syntax error is always fully parsed (not including any long-range dependencies, e.g. the final closing curly brace for a method definition, when the syntax error occurs in a statement in the middle of the method). When a syntax error is encountered during right-to-left parsing, the syntax error was not even previously encountered to the right of the syntax error. To the left of the syntax error, rules whose matches end before the start of the syntax error will still match their corresponding input; and just as with recursive descent parsing, rules whose matches start before the syntax error but overlap with the syntax error will fail to match. Consequently, pika parsers have \emph{optimal error recovery characteristics} without any further modification.

Syntax errors can be defined as regions of the input that are not spanned by matches of rules of interest. Recovering after a syntax error involves finding the next match in the memo table after the end of the syntax error for any grammar rule of interest: for example, a parser could skip over a syntax error to find the next complete function, statement, or expression in the input. This lookup requires $O(\mathrm{log} \, n)$ time in the length of the input if a skip list or balanced tree is used to store each row of the memo table.

\begin{figure}[ht]
    \centering
    \includegraphics[width=1\textwidth]{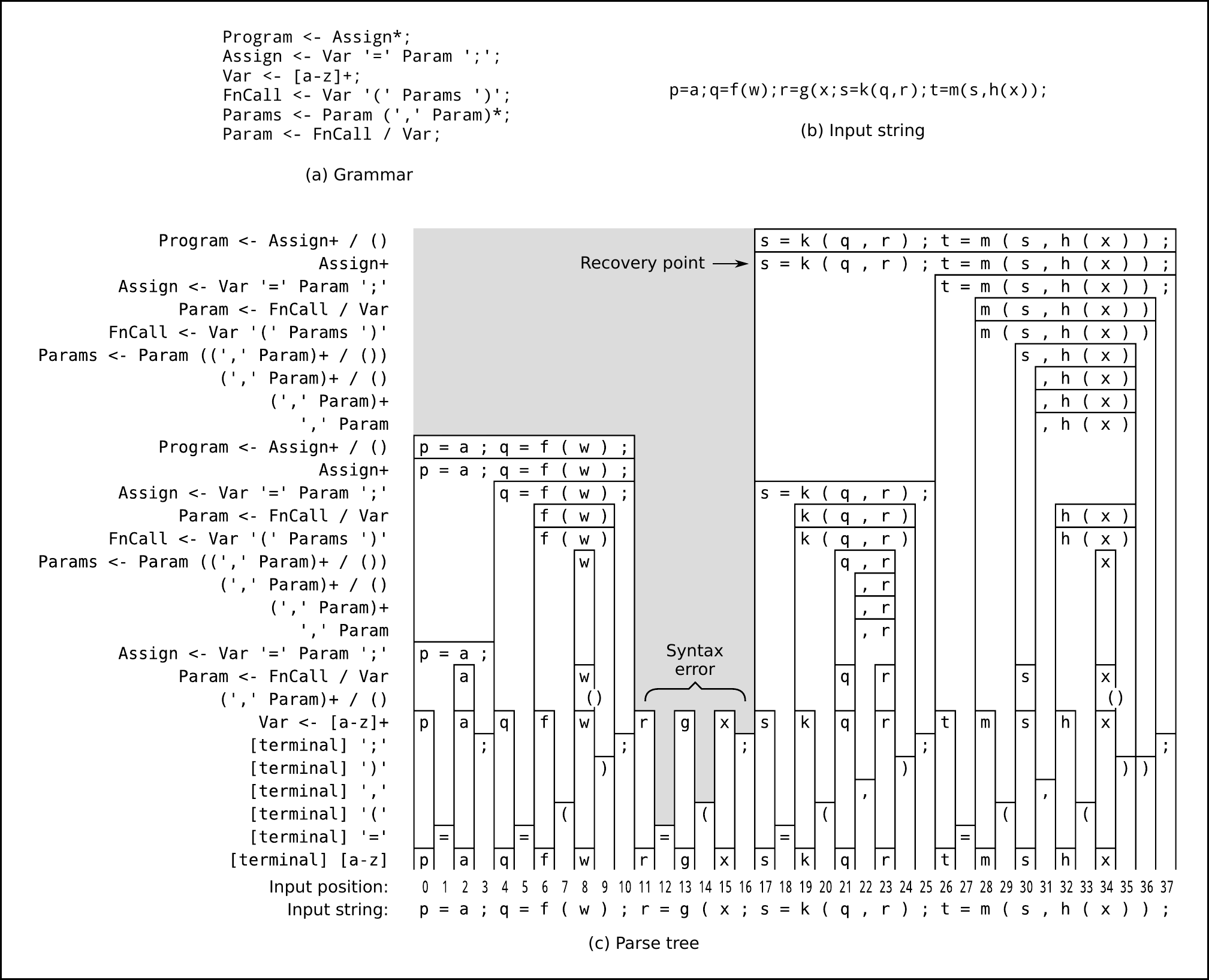}
    \caption{\label{fig:SyntaxError}Example of optimal recovery after a syntax error (a missing parenthesis after input position 15)}
    \Description[Optimal error recovery]{Example of optimal recovery after a syntax error}
\end{figure}

An example parse of an input containing a syntax error (a missing closing parenthesis after input position 15) is given in Fig.~\ref{fig:SyntaxError}. For this example, the beginning and end position of the syntax error are determined by finding regions of the input that are not spanned by a match of either the \texttt{Program} or \texttt{Assign} rule (any rules could be chosen for the purpose of identifying complete parses of rules of interest, and whatever is not spanned by those rules is treated as a syntax error). Using this method, the syntax error is found to be located between input positions 11 and 16 inclusive: the memo table contains terminal matches at these input positions, but no matches of either \texttt{Program} or \texttt{Assign}. Assuming that the intent is to recover at the next sequence of \texttt{Assign} rule matches after the syntax error, the parser can recover by finding the next match of \texttt{(Assign+)} in the memo table after the last character of the syntax error.

\subsection{\label{sec:precedence}Precedence parsing with PEG grammars}

A \emph{precedence-climbing grammar} tries to match a grammar rule at a given level of precedence, and, if that rule fails to match, defers or ``fails over'' to the next highest level of precedence (using a \texttt{First} clause, in the case of a PEG grammar). The highest precedence rule within a precedence hierarchy uses a \emph{precedence override pattern}, such as parentheses, to support multiple nested loops around the precedence hierarchy cycle.

A simple precedence-climbing grammar is shown in Listing~\ref{lst:grammarIncomplete}. The rule for each level of precedence in this example grammar has subclauses (corresponding to operands) that are references to rules at the next highest level of precedence. If the operands and/or the operator don't match, then the entire rule fails over to the next highest level of precedence. The root of the parse subtree for any fully-parsed expression will be a match of the lowest-precedence clause, in this case \texttt{E0}.

This simple example grammar can match nested unequal levels of precedence, e.g. \texttt{"1*2+3*4"}, but cannot match runs of equal precedence, e.g. \texttt{"1+2+3"}, since these cannot be parsed unambiguously without specifying the associativity of the addition operator. This grammar also cannot handle direct nesting of equal levels of precedence, e.g. \texttt{"-}\texttt{-4"}.

Since pika parsers can handle left recursion directly, these issues can be fixed by modifying the grammar into the form shown in Listing~\ref{lst:grammarFixed}, allowing unary minus and parentheses to self-nest, and resolving ambiguity in runs of equal precedence by rewriting rules into left-recursive form, which generates a parse tree with left-associative structure. Similarly, a right-associative rule \texttt{Y1} employing a binary operator \texttt{OP} would move the self-recursive subclause to the right side, yielding the right-recursive form \texttt{Y1 <- (Y2 OP Y1) / Y2}, which would generate a parse tree with right-associative structure.

The reference pika parser (Section~\ref{sec:software}) can automatically rewrite grammars into the form shown in Listing~\ref{lst:grammarFixed} if supplied with the syntax shown in Listing~\ref{lst:grammarRefParserSyntax}, which can include precedence level and an optional ``\texttt{,L}'' or ``\texttt{,R}'' associativity-specifying parameter within square brackets after the rule name.

\section{\label{sec:Performance}Performance and benchmarking}

Ignoring cycles in the grammar, packrat parsing would appear to take $O(n|G|)$ time in the worst case for input length $n$ and grammar size $|G|$, since there are $n$ columns and $|G|$ rows in the memo table. However, a very deep imbalanced parse tree tends towards the structure of a linked list, therefore both the worst case depth of the parse tree and the number of nodes in the parse tree are linearly correlated with $n$, and the actual worst case performance of packrat parsing is actually $O(n)$ no matter the structure or size of the grammar, assuming full memoization of recursive calls.

For pika parsing, there is a certain amount of work is wasted due to spurious matches (i.e. matches that will not end up in the final parse tree) caused by working bottom-up from terminals without taking parsing context into account. Initialization of bottom-up parsing requires matching $|T|$ terminals against $n$ input positions, requiring $O(n|T|)$ time. Spurious matches of terminals against the input (e.g. matching a word inside a comment as an identifier rather than a sequence of comment characters) may subsequently trigger higher-level spurious matches, up to some worst case height $h$ where no more higher-level structure can be spuriously matched. (Under most circumstances, with the exception of special cases like commented-out code, spurious matches usually do not cascade far up the grammar hierarchy, because random input does not typically match large parts of the grammar with high likelihood). The problem of spurious matches is exacerbated when grammars reuse subclauses between many rules: this can lead to the shared subclauses having multiple seed parent clauses, which can trigger an upwards-directed tree of spurious matches, effectively further increasing the overhead $h$ of spurious matches. The work required to find the correct matches and build the parse tree is $O(n)$, as for top-down parsing. This gives an overall runtime of $O(n|T| + nh + n)$. The number of terminals is constant for a given grammar, and it is a reasonable simplifying assumption that the worst-case ratio of spurious matches to correct matches is constant for the grammar. This reduces both the time and the memory complexity of pika parsing \emph{for a specific grammar} to $O(kn)$, where $k$ is some constant overhead factor of pika parsing relative to packrat parsing for the grammar, due to spurious matches. The factor $k$ is a complex function of not only the size and complexity of the grammar, but also of the distribution of all inputs that may be reasonably parsed -- however, for a \emph{fixed} grammar and a distribution of inputs, $k$ can be assumed to be constant, i.e. pika parsing is $O(n)$ in the length of the input.

This linear scaling property of pika parsing was measured and demonstrated to hold for simple and complex grammars.

\subsection{Benchmarking setup}

The reference pika parser (Section~\ref{sec:software}) was benchmarked against the two widely-used JVM parsing libraries: Parboiled2 version 1.3.1~\citep{parboiled2} and ANTLR4 version 4.8-1~\citep{antlr4}. ANTLR4 sees particularly widespread usage, because its polyglot code generator allows the library to generate parsers for not only standard JVM languages, but also C\#, Python, JavaScript, C++, Swift, Go and PHP.

Parboiled2 is a Scala-based (partially memoizing) packrat parsing library for PEG grammars, and does not handle left recursion. ANTLR4 is a Java-based memoizing ALL(*) parser generator, and handles direct left recursion but does not handle indirect left recursion (ANTLR3 handled neither form of left recursion). Both Parboiled2 and ANTLR4 are considered to have best-in-class performance, although both require careful grammar design to avoid superlinear scaling behaviors in the worst case.

Two grammars were chosen for benchmarking:

\begin{itemize}
    \item The \emph{expression grammar} shown in Listing~\ref{lst:grammarIncomplete}.
    \item The \emph{Java grammar}, an implementation of the Java language specification. For the pika parser and Parboiled2, a PEG grammar implementing the Java 6 language specification was obtained from the Parboiled2 software distribution. ANTLR4 is not a PEG parser, so this same grammar could not be used for ANTLR4, but a comparable grammar, implementing the Java 8 language specification and provided in the ANTLR4 software distribution, was used instead.
\end{itemize}

Note that any grammar for the Java language specification is complex, widely employing deep precedence climbing hierarchies throughout the grammar, which can result in parse trees dozens to hundreds of nodes deep. This grammar is arguably within the same order of magnitude of complexity as any of the most complex grammars in use for real-world programming languages. The Parboiled2 PEG grammar consists of 669 clauses, of which 143 are terminals. Of the 669 clauses in the grammar, 100 clauses had at least 2 seed parent clauses; 50 clauses had at least 4 seed parent clauses; 4 clauses had at least 10 seed parent clauses; and the clause with the highest number of seed parent clauses, the clause matching an identifier, \texttt{(!Keyword Letter LetterOrDigit* Spacing)}, had 15 seed parent clauses. (The empty string matching clause, \texttt{Nothing}, had 88 seed parent clauses, but this is handled differently by the pika parser to reduce the number of memoized zero-length matches -- Section~\ref{sec:zerolen}.)

For each grammar type, a corresponding input dataset was generated.

\begin{itemize}
    \item The expression grammar is a simple precedence-climbing grammar that does not include direct or indirect left recursion (Parboiled2 and ANTLR4 cannot handle one or both forms of left recursion). This requires inputs to be fully parenthesized, so that runs of two or more operators of equal precedence do not occur in the input. In total, 2500 random arithmetic expressions were generated, consisting of integers, binary operators, unary negation and parentheses, nested up to 25 levels deep. This resulted in arithmetic expression strings with a minimum length of 5 characters, an average length of 140k characters, and a maximum length of 4.8M characters.
    \item For the Java grammar, 5372 Java 6 source files were sourced from an old Java 6 version of the Spring framework, with an average size of 3.8kiB and a maximum size of 76kiB. Comments were stripped from the input files (to focus on the performance of parsing the structural elements of each source file), but string constants (which can also cause spurious matches for the pika parser) were left in place. (A more highly optimized pika parser would reduce spurious matches with a lex preprocessing step, see Section~\ref{sec:lex} -- although this was not attempted for this benchmark.)
\end{itemize}

Each input from each dataset was provided to each of the three parsers, and parsed with the appropriate grammar. All benchmarks were run single-threaded with a Ryzen-1800X CPU using JRE version 13.0.2.

\subsection{The scaling characteristics of pika parsing}

\begin{figure*}[ht]
    \centering
    \begin{subfigure}[b]{0.48\textwidth}
        \centering
        \includegraphics[width=\textwidth]{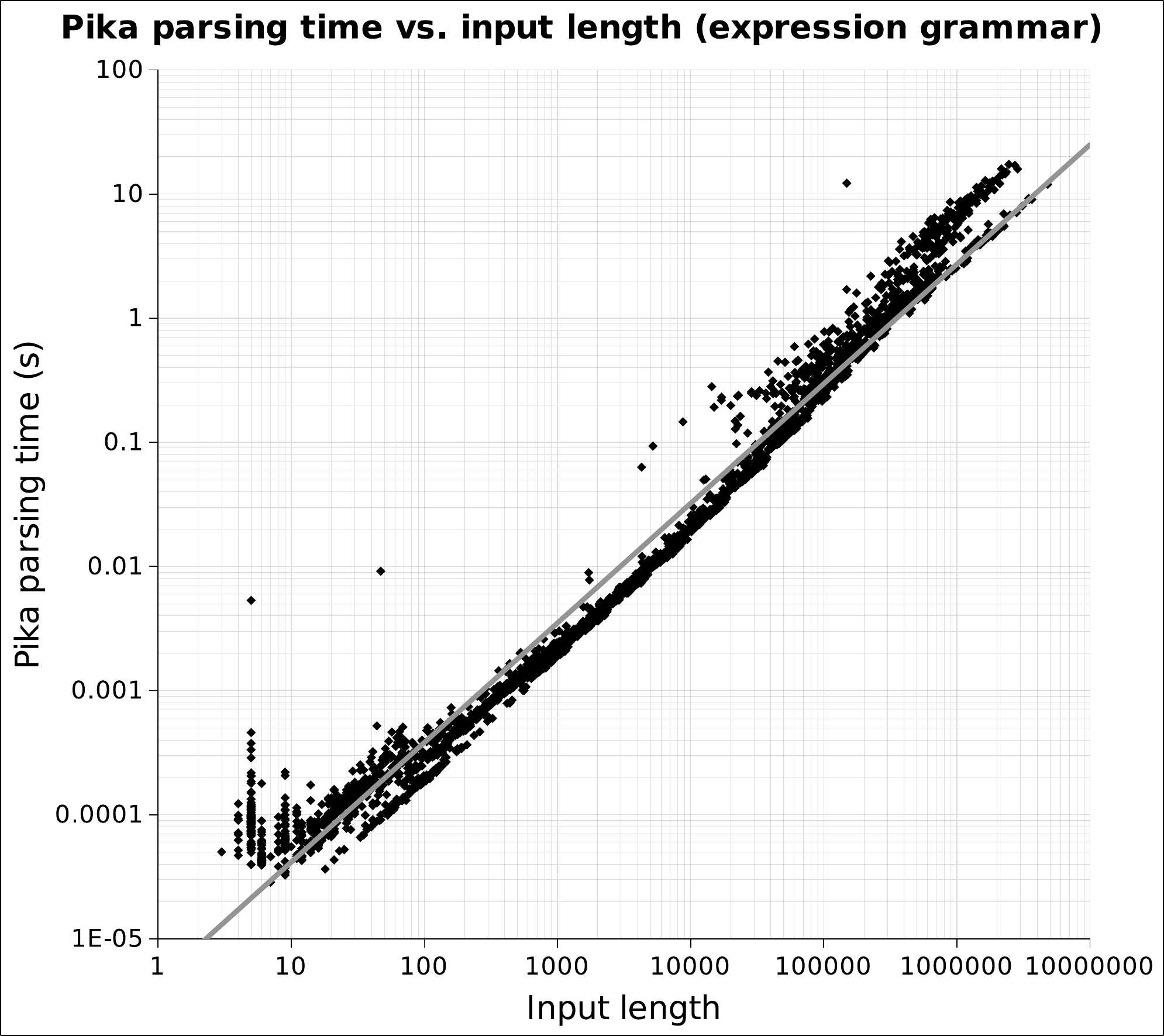}
        \caption{}
        \label{chart:Pika-vs-len-expr}
    \end{subfigure}
    \hfill
    \begin{subfigure}[b]{0.48\textwidth}
        \centering
        \includegraphics[width=\textwidth]{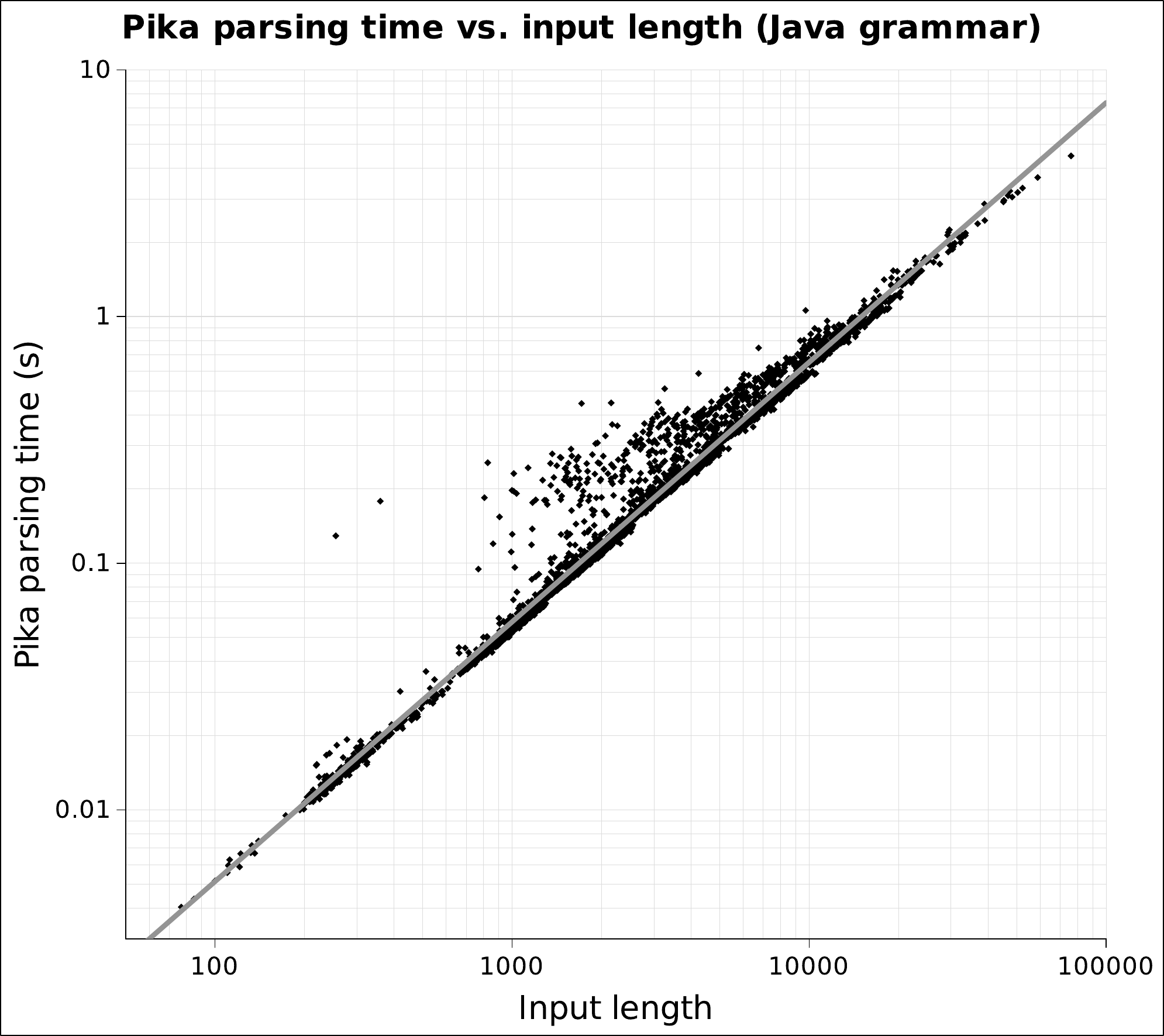}
        \caption{}
        \label{chart:Pika-vs-len-Java}
    \end{subfigure}
    \caption{\label{chart:Pika-vs-len}Parsing time for the pika parser as a function of input length.}
    \Description[Parsing time for the pika parser, as a function of input length]{Parsing time for the pika parser, as a function of input length}
\end{figure*}

As depicted in Fig.~\ref{chart:Pika-vs-len} and summarized in Table~\ref{tab:Benchmarks-vs-len}, pika parsing was measured as scaling linearly in the length of the input. Note that a straight line $(\mathrm{ln} \, y = m \, \mathrm{ln} \, x + c)$ in a log-log plot corresponds to the polynomial $(y = e^c \, x^m)$, so an apparent linear correlation in a log-log plot does not imply a linear correlation if the same data were plotted with linear axes unless $(m = 1)$.

Parsing cannot be sub-linear, since all input must be consumed, so the fact that the smaller of the two scaling powers of the regression line (0.963) fell further below 1.0 than the larger scaling power (1.05) fell above 1.0 indicates that both these grammars are within reasonable error bounds of linear scaling. The $R^2$ values for the regression lines are both very close to 1.0, indicating that the linear correlation is strong.

Notably, this linear relationship between input length and parsing time held for both the simple expression grammar and the significantly more complex Java grammar, across several orders of magnitude of input length. This is a very favorable trait for a parsing algorithm.

The pika parsing speed for the expression grammar (in characters of input processed per second) is approximately 8.9 times higher than the parsing speed for the expression grammar: $1/0.0000045 = 222,000$ vs. $1/0.000040 = 25,000$ characters per second respectively. For reference, the Java grammar consists of over 61 times as many clauses as the expression grammar, 6.85 times greater than the difference in parsing speed, so the impact of increasing the size of the grammar on the parsing speed appears to be sublinear. This is further reinforced by the fact that the larger Java grammar is also significantly more structurally complex than the expression grammar, and shares subclauses widely, leading to more seed parent clauses per clause, which will trigger more spurious matches than simply increasing the size of the grammar alone.

\begin{table*}[htbp]
    \centering
    {\small
    \begin{tabular}{cccccc}\toprule
    
    \textbf{Figure} & \textbf{Grammar} &  \textbf{y} & \textbf{x} & \textbf{Regression} & $\bm{R^2}$\\\midrule
    
    \ref{chart:Pika-vs-len-expr} & expression & Pika parsing time & input length & $y = 0.0000045 \, x^{0.963}$ & 0.974 \\
    \ref{chart:Pika-vs-len-Java} & Java & Pika parsing time & input length & $y = 0.000040 \, x^{1.05}$ & 0.975 \\\bottomrule
    \end{tabular}
    }
    \vspace{2mm}
    \caption{\label{tab:Benchmarks-vs-len}Benchmarks of pika parsing time vs. input length.}
    \Description[Benchmarks of pika parsing time vs. input length]{Benchmarks of pika parsing time vs. input length}
\end{table*}

\subsection{Parsing time for ANTLR4 and Parboiled2 vs. the reference pika parser: expression grammar}

\begin{figure*}[ht]
    \centering
    \begin{subfigure}[b]{0.48\textwidth}
        \centering
        \includegraphics[width=\textwidth]{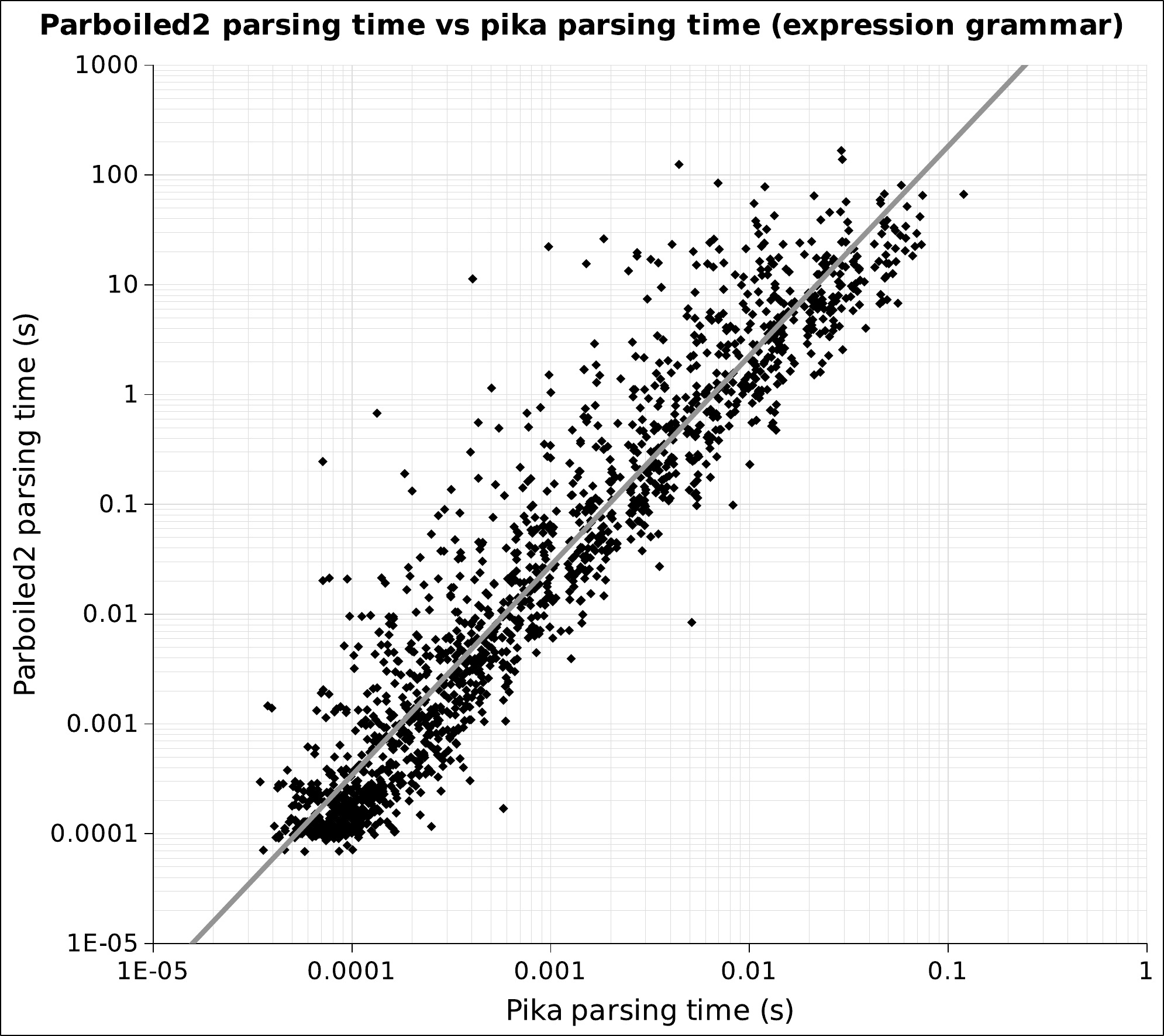}
        \caption{}
        \label{chart:Parboiled-vs-pika-expr}
    \end{subfigure}
    \hfill
    \begin{subfigure}[b]{0.48\textwidth}
        \centering
        \includegraphics[width=\textwidth]{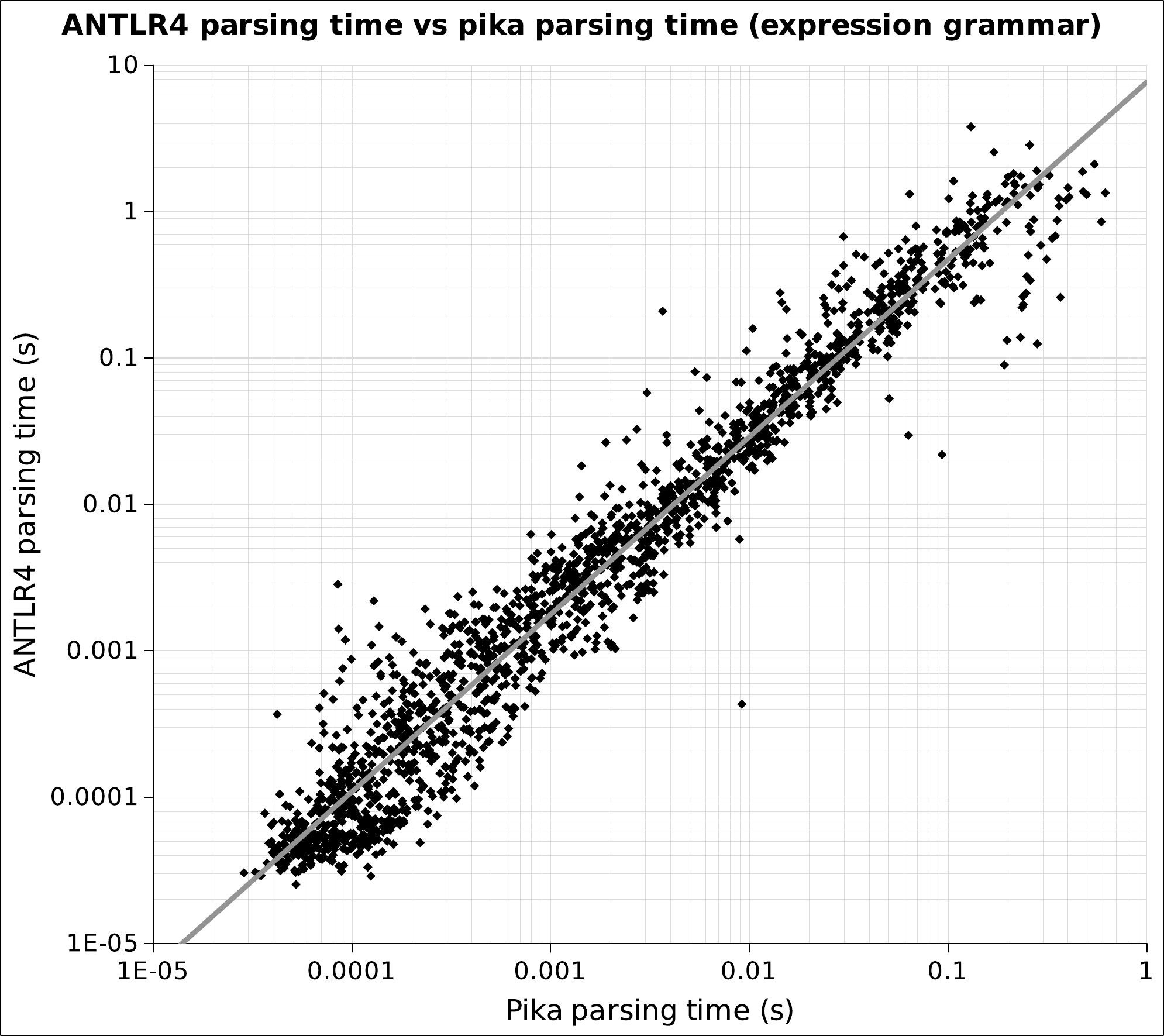}
        \caption{}
        \label{chart:ANTLR-vs-pika-expr}
    \end{subfigure}
    \\
    \begin{subfigure}[b]{0.48\textwidth}
        \centering
        \includegraphics[width=\textwidth]{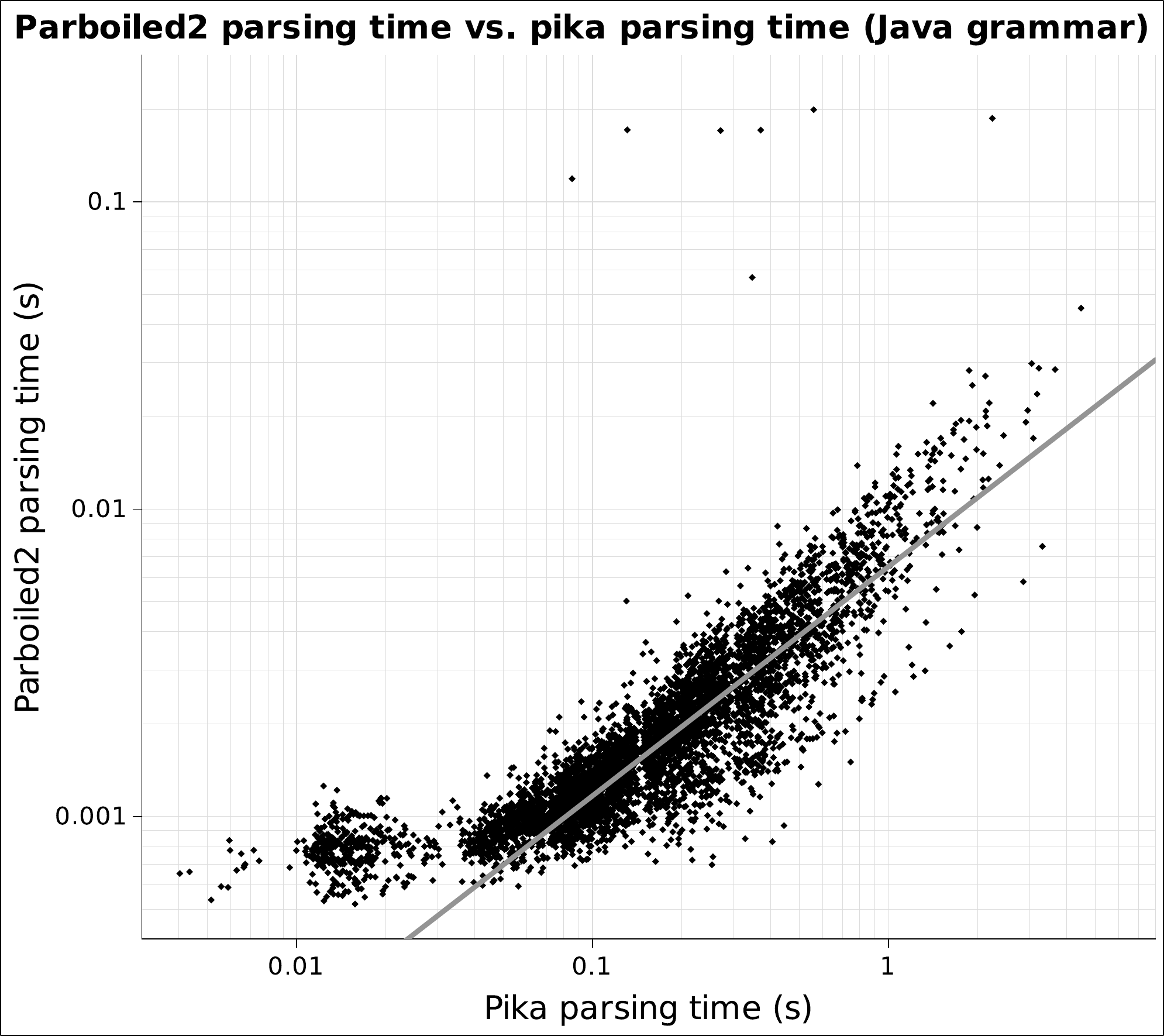}
        \caption{}
        \label{chart:Parboiled-vs-pika-Java}
    \end{subfigure}
    \hfill
    \begin{subfigure}[b]{0.48\textwidth}
        \centering
        \includegraphics[width=\textwidth]{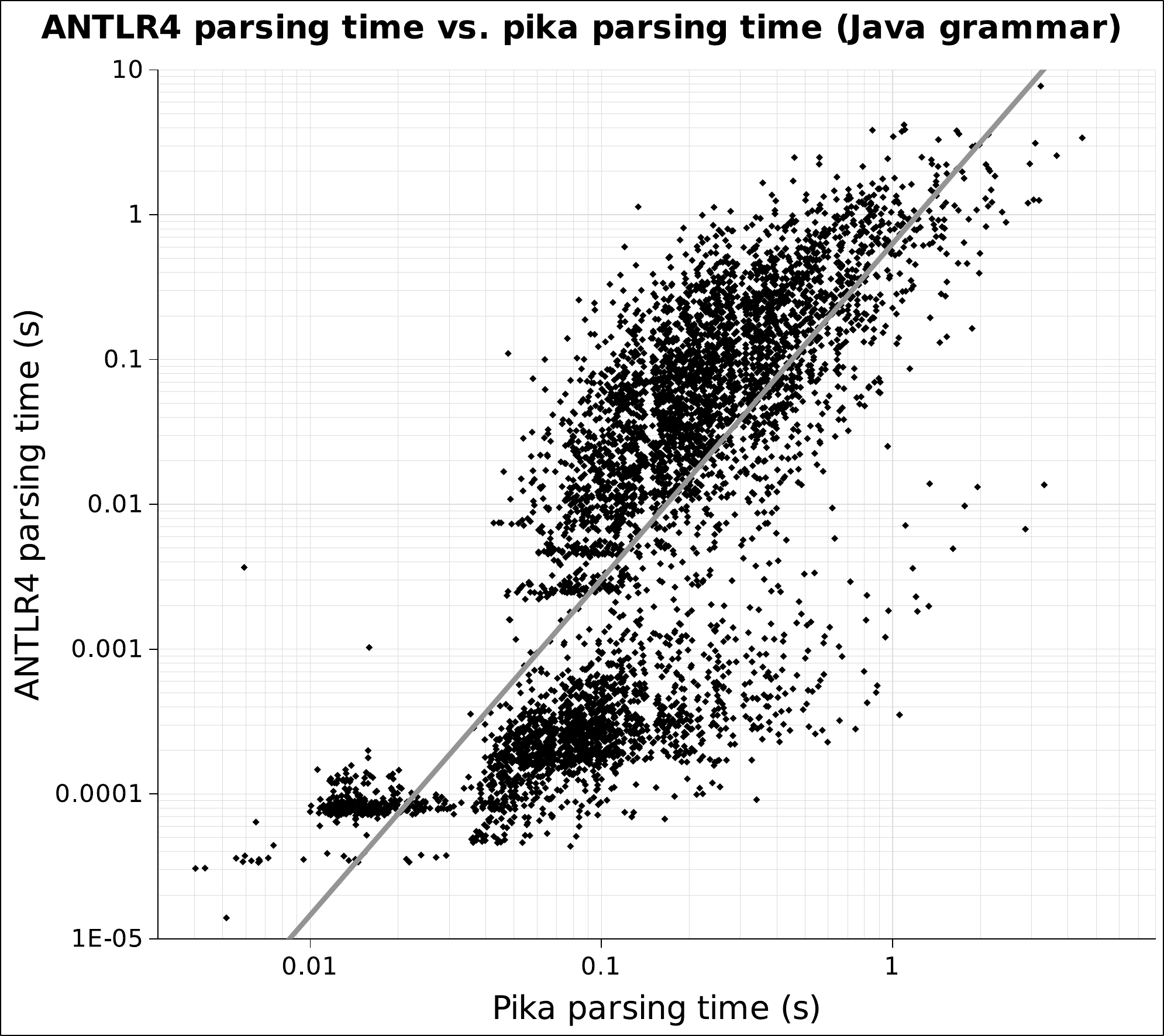}
        \caption{}
        \label{chart:ANTLR-vs-pika-Java}
    \end{subfigure}
    \caption{\label{chart:Others-vs-pika}Parsing time for the Parboiled2 and ANTLR4 parsers vs. the pika parser, for the expression grammar and the Java grammar.}
    \Description[Parsing time for the Parboiled2 and ANTLR4 parsers vs. the pika parser]{Parsing time for the Parboiled2 and ANTLR4 parsers vs. the pika parser}
\end{figure*}

As depicted in Fig.~\ref{chart:Others-vs-pika} and summarized in Table~\ref{tab:Benchmarks-vs-pika}, the performance of the pika parser relative to other parsers depends greatly upon the grammar in use, and the length of the input.

For the expression grammar, the pika parser was roughly the same speed as Parboiled2 for small expressions, but the pika parser was 1000 times faster than Parboiled2 for larger expressions. In fact, Parboiled2 degraded nearly quadratically as a function of the length of the input for this grammar (as seen in the power term of the regression line, $x^{1.909}$, where $x$ is the parsing time for the pika parser, which is a linear function of the length of the input).

The pika parser was also roughly the same speed as ANTLR4 for small expressions, but the pika parser was 10 times faster than ANTLR4 for large expressions, and ANTLR4 also degraded superlinearly in performance as a function of input length ($x^{1.212}$).

\subsection{Parsing time for ANTLR4 and Parboiled2 vs. the reference pika parser: Java grammar}

For the Java grammar, Parboiled2 was estimated by regression to scale sublinearly in the length of the input ($x^{0.746}$); however, this is impossible, since all input must be consumed by any parser. The regression power being below 1.0 merely demonstrates that Parboiled2 was superlinearly inefficient for small inputs, and became more efficient as the input size increased: for very long inputs, the performance of Parboiled2 must trend to at least linear. It can be seen then that the pika parser is approximately 100 times slower than Parboiled2 for moderate to long input lengths, whereas for shorter inputs, the pika parser was 10 times slower than Parboiled2. This demonstrates the overhead of spurious matches incurred by bottom-up matching with a large and complex grammar.

ANTLR4 is harder to analyze, because it exhibited an inconsistent (and multimodal) distribution of parsing times relative to the pika parser. For shorter inputs, the pika parser was up to 1000, 100 or 10 times slower than ANTLR4, whereas for the longest inputs, the pika parser and the ANTLR parser ran at roughly the same speed. Despite the slower speed for shorter inputs, the pika parser has a significantly more predictable parsing time as a function of input length than ANTLR4, and also has better scaling characteristics: parsing time for ANTLR4 scales between quadratically and cubically in the length of the input, as can be seen from the power of the regression line, $x^{2.32}$ (note however that the performance characteristics of ANTLR4 are quite unpredictable, with $R^2 = 0.6$, so the regression line is only a rough estimate of the scaling properties of ANTLR4 relative to the pika parser).

\begin{table*}[htbp]
    \centering
    {\small
    \begin{tabular}{cccccc}\toprule
    
    \textbf{Figure} & \textbf{Grammar} &  \textbf{y} & \textbf{x} & \textbf{Regression} & $\bm{R^2}$\\\midrule

    \ref{chart:Parboiled-vs-pika-expr} & expression & Parboiled2 parsing time & Pika parsing time & $y = 14804 \, x^{1.909}$ & 0.897 \\
    \ref{chart:ANTLR-vs-pika-expr} & expression & ANTLR4 parsing time & Pika parsing time & $y = 7.682 \, x^{1.212}$ & 0.957 \\\midrule

    \ref{chart:Parboiled-vs-pika-Java} & Java & Parboiled2 parsing time & Pika parsing time & $y = 0.0065 \, x^{0.746}$ & 0.614 \\
    \ref{chart:ANTLR-vs-pika-Java} & Java & ANTLR4 parsing time & Pika parsing time & $y = 0.63 \, x^{2.32}$ & 0.600 \\\bottomrule
    \end{tabular}
    }
    \vspace{2mm}
    \caption{\label{tab:Benchmarks-vs-pika}Benchmarks of parsing time for the Parboiled2 and ANTLR4 parsers vs. the pika parser.}
    \Description[Benchmarks of parsing time for the Parboiled2 and ANTLR4 parsers vs. the pika parser]{Benchmarks of parsing time for the Parboiled2 and ANTLR4 parsers vs. the pika parser}
\end{table*}

\subsection{Performance discussion}

The reference pika parser was zero to three orders of magnitude faster than Parboiled2 and zero to one order of magnitude faster than ANTLR4 for inputs conforming to the expression grammar, with every indication that the performance gap would widen further for even larger inputs. However, the pika parser was one to two orders of magnitude slower than Parboiled2 and zero to three orders of magnitude slower than ANTLR4 for inputs conforming to the Java grammar, with the performance gap narrowing as input length increased for the ANTLR4 case (i.e. the large constant performance overhead of pika parsing would eventually be dominated by the super-quadratic scaling behavior of ANTLR4). Parboiled2 appeared to maintain linear performance for large inputs. This indicates that the pika parser may not be the right tool for every parsing task, depending on the grammar, the expected range of input sizes, and the degree to which parsing performance is important for a given application.

However, the parse time for the pika parser consistently demonstrated a strongly linear relationship to the length of the input, for both the simple expression grammar and the complex Java grammar, in spite of large differences in size and structure between the two grammars. The strongly linear performance characteristics measured for the pika parser increase the attractiveness of this parsing algorithm for real-world usage, because the performance is very predictable. The widely-used Parboiled2 and ANTLR4 libraries both claim linear scaling under typical usage; however, each of these libraries exhibited quadratic or even super-quadratic performance degradation for one of the two grammars, and ANTLR4 in particular appears to suffer from multi-modal performance degradation behavior (Fig.~\ref{chart:ANTLR-vs-pika-Java}).

Some effort was made to understand cherry-picked cases where Parboiled2 and ANTLR4 were both able to parse a Java source file one or more orders of magnitude more quickly than the pika parser. In each of these cases, the source files were long, with very shallow syntactic structure (e.g. the files consisted of mostly constant static fields with literal initializer values). Deep grammatical structures slow down ANTLR4 in particular, so it is significantly more efficient when parsing only shallow grammatical structures. Parboiled2 used numerous tricks to speed up parsing, for example not memoizing the bottom few tiers of the parse tree (e.g. not memoizing identifiers character-by-character). These sorts of optimizations were not applied to the reference pika parser, further separating the performance of the pika parser from the other parsers.

The most significant mechanism for speeding up the pika parser is to reduce the number of spurious matches through the use of a lex preprocessing step (Section~\ref{sec:lex}), in particular representing quoted strings and comments as a single token. This was not attempted for the Java grammar in these benchmarks, because complex work would be required for such a large grammar to ensure that lexing the input did not change the recognized language.

Many grammars are nowhere near as complex as the grammar for the Java language specification, and the pika parser appears to perform efficiently for grammars that are not as large or complex. Even for large or complex grammars, the overhead of pika parsing relative to packrat parsing may be acceptable, particularly where the benefits of direct handling of left recursion and optimal error recovery are worth the constant performance penalty, and/or where linear scaling in the input length must be guaranteed.

However, whether or not the pika parsing algorithm is usable for any particular real-world parsing usecases, the algorithm should still be of interest from a theoretical perspective, since the inversion of recursive descent parsing into dynamic programming form raises interesting possibilities for other traditionally top-down recursive algorithms.

\section{Implementation details}

The key classes of the pika parser are as follows. The code shown is simplified to illustrate only the most important aspects of the parsing algorithm and data structures (e.g. field initialization steps and trivial assignment constructors are usually omitted; optimizations are not shown; etc.). Listings are given in valid Java notation for precise semantic clarity compared to an invented pseudocode\footnote{Java is a reasonable choice for code snippets in published papers due to its popularity, its low degree of implicitness, and its general readability even by most non-Java programmers. The use of \emph{ad hoc} invented pseudocodes in the majority of computer science papers over many years has contributed to a crisis of reproducibility, due to ambiguities in pseudocode notation, important omitted details in pseudocode snippets, and outright errors accidentally introduced when rewriting a working implementation into pseudocode. Often these issues only become apparent when a reader attempts to convert the pseudocode in a paper into an implementation in a specific language.}; however, Java-specific features and classes were avoided wherever possible, so that the code is mostly generic, and should be readable to anyone with an understanding of C++-derived object oriented languages. See the reference parser for the full implementation (Section~\ref{sec:software}).

\subsection{\label{sec:MemoKey}The \texttt{MemoKey} class}

The \texttt{MemoKey} class (Listing~\ref{lst:MemoKey}) has two fields, \texttt{clause} and \texttt{startPos}, respectively corresponding to the row and column of an entry within the memo table\footnote{The \texttt{hashCode} and \texttt{equals} methods, required by Java to implement the key equality contract so that \texttt{MemoKey} can be used as a hashtable key, and the trivial assignment constructor, are not shown.}.

\subsection{\label{sec:Grammar}The \texttt{Grammar} class}

The \texttt{Grammar} class (Listing~\ref{lst:Grammar}) has a field \texttt{allClauses}, containing all unique clauses and subclauses in the grammar in bottom-up topological order (Section~\ref{sec:topoOrder}), and one primary method, \texttt{parse}, which contains the main parsing loop that matches all relevant clauses for each start position using a priority queue (Section~\ref{sec:handlingCycles}), memoizing any matches that are found. Surprisingly, this \texttt{parse} method is the entire pika parsing algorithm, at the highest level. However, \texttt{Clause.match} (Section~\ref{sec:Clause}) and \texttt{MemoTable.addMatch} (Section~\ref{sec:MemoTable}) still need to be detailed.

\subsection{\label{sec:Clause}The \texttt{Clause} class and its subclasses}

The \texttt{Clause} class (Listing~\ref{lst:Clause}) represents a PEG operator or clause, and is linked to any subclauses via its \texttt{subClauses} field. This class has a subclass \texttt{Terminal}, which is the superclass of all terminal clause implementations, whereas nonterminals extend \texttt{Clause} directly.

Each subclass of \texttt{Clause} subclass must implement the following three methods:

\begin{itemize}
    \item The \texttt{determineWhetherCanMatchZeroChars} method sets the \texttt{canMatchZeroChars} field to true if the clause can match while consuming zero characters (see Section~\ref{sec:zerolen}). This value may depend upon the \texttt{canMatchZeroChars} field of each of the subclauses, therefore this method must be called in bottom-up topological clause order.
    \item The \texttt{addAsSeedParentClause} method adds a clause to its subclauses' \texttt{seedParentClauses} list, if the clause and the subclause could be matched at the same starting position.
    \item The \texttt{match} method checks whether all the necessary subclauses of the clause match in the memo table, and if so, returns a new \texttt{Match} object, otherwise returns \texttt{null}.
\end{itemize}

The \texttt{Char} class (Listing~\ref{lst:Char}), which extends \texttt{Terminal}, which extends \texttt{Clause}, is a simple example of a terminal clause that can match a single specific character in the input. Terminals directly check whether they match the input string, rather than looking in the memo table as with nonterminal clauses, so they ignore the \texttt{memoTable} parameter. (Other methods are not shown for simplicity.)

The \texttt{Seq} class (Listing~\ref{lst:Seq}), which extends \texttt{Clause}, implements the \texttt{Seq} PEG operator. In order to reduce the number of zero-length matches added to the memo table, the \texttt{determineWhetherCanMatchZeroChars} method marks the \texttt{Seq} clause as able to match zero characters if all subclauses are able to match zero characters, and the \texttt{addAsSeedParentClause} method adds the \texttt{Seq} clause to the \texttt{seedParentClauses} field in each of its subclauses, up to and including the first subclause that consumes at least one character in any successful match. The \texttt{match} method returns a new \texttt{Match} object if all subclauses of the \texttt{Seq} clause are found to match in consecutive order, otherwise the method returns \texttt{null}. Any returned \texttt{Match} object may form one node in the final parse tree.

The other subclasses of \texttt{Clause} (implementing the other PEG operator types) can be found in the reference parser (Section~\ref{sec:software}).

\subsection{\label{sec:Match}The \texttt{Match} class}

The \texttt{Match} class (Listing~\ref{lst:Match}) is used to store matches (i.e. parse tree nodes) and their lengths (i.e. the number of characters consumed by the match) in memo entries, and to link matches to their subclause matches (i.e. to link parse tree nodes to their child nodes). For \texttt{First} clauses, the index of the matching subclause is also stored in the \texttt{firstMatchingSubClauseIdx} field of the \texttt{Match} object, so that a match of an earlier subclause can take priority over any match of later subclauses, as required by the semantics of the \texttt{First} PEG operator.

Two matches can be compared using the \texttt{isBetterThan} method to determine whether a newer match improves upon an earlier match in the same memo entry, as defined in Section~\ref{sec:leftRecursionTermination}.

\subsection{\label{sec:MemoTable}The \texttt{MemoTable} class}

The \texttt{MemoTable} class (Listing~\ref{lst:MemoTable}) is a wrapper for the memo table hashmap, stored in the field \texttt{memoTable}.

The \texttt{lookUpBestMatch} method looks up a memo entry by memo key, i.e. by clause and start position, returning any \texttt{Match} object stored in the memo entry corresponding to this memo key.

However, one tweak is necessary for \texttt{NotFollowedBy} subclauses: if there is no memo table entry for a \texttt{NotFollowedBy} subclause, then it is impossible to determine whether the \texttt{NotFollowedBy} subclause's own subclause matched, or was not previously evaluated for a match. In other words, the inversion logic of the \texttt{NotFollowedBy} PEG operator does not work directly with the upwards-expansion of the DP wavefront, since seed parent clauses are only typically scheduled for matching when a subclause actually matches; however, seed parent clauses are also not scheduled for matching if the subclause is itself never tested for match. Therefore, when there is no known match in the memo table entry for the subclause of a \texttt{NotFollowedBy} clause, the subclause must be evaluated top-down to determine whether it matches in terms of whether or not its own subclause matches. 

There is also one optimization made to \texttt{lookUpBestMatch} with the goal of reducing the number of memoized zero-length matches (Section~\ref{sec:zerolen}): if there is no memoized match for a given memo key, but the clause in the memo key can match zero characters (e.g. \texttt{(X?)}), then a zero-length match must be returned. (This is required since \texttt{Nothing} is excluded from the list of terminals that are matched bottom-up, for efficiency, since this clause matches at all input positions.)

The \texttt{addMatch} method checks whether the provided \texttt{match} parameter is not \texttt{null} (indicating a mismatch), and if not, looks up the current best match for the memo key. If there is no current best match, or the current best match is not as good a match as the newer match, then the memo entry is updated with the new match. If the memo entry was updated, the seed parent clause(s) of the matching clause are added to the priority queue to be scheduled for matching at the same start position.

As an optimization, again for reducing the number of memoized zero-length matches, even if the memo entry was not updated or if the match was \texttt{null}, and if the parent clause can match zero characters, then the seed parent clause is still scheduled for matching (since the ability of the parent clause to match zero characters means the parent will always match, whether or not the subclause matches).

\section{Optimizations}

\subsection{\label{sec:zerolen}Reducing the size of the memo table by avoiding memoizing zero-length matches}

Seeding bottom-up matching from all terminals, including \texttt{Nothing} (which matches the empty string $\epsilon$ in every character position, consuming zero characters) is inefficient, because many entries will be created in the memo table for zero-length matches that will not be linked into the final parse tree.

\texttt{Nothing} is the simplest example of a clause that can match zero characters, but there are numerous other examples of PEG clauses that can match zero characters, for example any instance of \texttt{Optional}, such as \texttt{(X?)}; any instance of \texttt{ZeroOrMore}, such as \texttt{(X*)}; instances of \texttt{First} where any subclause (specifically the last subclause\footnote{If a \texttt{First} clause has any subclause that can match zero characters, then all subsequent subclauses of the \texttt{First} clause will be ignored, because the subclause that can match zero characters will always match.}) can match zero characters, such as \texttt{(X / Y / Z?)}; and instances of \texttt{Seq} where all subclauses can match zero characters (causing the whole \texttt{Seq} clause to match zero characters), such as \texttt{(X? Y* Z?)}. Zero-length matches can cascade up the grammar from these terminals towards the root, adding match entries to the memo table at every input position for multiple levels of the grammar. This can be mitigated by not seeding upwards propagation of the DP wavefront from the \texttt{Nothing} match in every character position -- however, then clauses that can match zero characters must be handled specially.

It is safe to assume that no clause will never have \texttt{Nothing} as its first subclause, since this would be useless for any type of parent clause. If \texttt{Nothing} is disallowed in the first subclause position, then it is unnecessary to trigger upwards expansion of the DP wavefront by seeding the memo table with zero-length $\epsilon$ matches at every input position during initialization, earlier non-\texttt{Nothing} subclauses of a \texttt{Seq} clause can fill the role of triggering the parent clauses.

However, when \texttt{Nothing} is never matched as a terminal when seeding upwards expansion of the DP wavefront, other subclause types that can match zero characters must be treated specially, in particular in the case of \texttt{Seq} clauses, which must be triggered as seed parent clauses by all subclauses that may match in the same start position, which means all subclauses up to and including the first subclause that must match one character or more (Listings~\ref{lst:Seq}, \ref{lst:MemoTable}). This mechanism for minimizing the memoization of zero-length matches requires the \texttt{Clause.canMatchZeroChars} field to be set for each clause during initialization, in bottom-up topological order.

With each of these changes in place, parsing is able to complete successfully and correctly while greatly reducing the number of memoized zero-length matches. The full implementation of these optimization steps can be seen in the reference parser (Section~\ref{sec:software}).

\subsection{Reducing the size of the memo table by rewriting \texttt{OneOrMore} into right-recursive form}

A na\"{\i}ve implementation of the \texttt{OneOrMore} PEG operator is iterative, matching as many instances of the operator's single subclause as possible, left-to-right, and assembling all subclause matches at each match position into an array, which is stored in the \texttt{subClauseMatches} field of the resulting \texttt{Match} object. However, when parsing the input from right to left, a rule like \texttt{(Ident <- [a-z]+)} adds $O(m^2)$ match nodes to the memo table as a function of the maximum number of subclause matches of a \texttt{OneOrMore} clause, $m$. For example, given the grammar clause \texttt{([a-z]+)} and the input string \texttt{"hello"}, this rule stores the following matches in the memo table: \texttt{[ [h,e,l,l,o], [e,l,l,o], [l,l,o], [l,o], [o] ]}.

This can be resolved by rewriting \texttt{OneOrMore} clauses into right-recursive form, which turns the parse tree fragment that matches the \texttt{OneOrMore} clause and its subclauses into a linked list. For example, the \texttt{OneOrMore} rule \texttt{(X <- Y+)} can be rewritten as \texttt{(X <- Y X?)}. Correspondingly, a \texttt{ZeroOrMore} rule \texttt{(X <- Y*)} can be rewritten as \texttt{(X <- (Y X?)?)}. The effect of this rewriting pattern is to store successively shorter suffix matches as list tails, rather than each suffix of subclause matches being duplicated for each match position. Each match of \texttt{X} in the rewritten form consists of either one or two subclause matches: a single match of subclause \texttt{Y}, optionally followed by a nested match of \texttt{X} representing the tail of the list, comprised of successive matches of \texttt{Y}. With this modification, the number of subclause matches created and added to the memo table becomes linear (i.e. $O(m)$) in the maximum number of subclause matches of a \texttt{OneOrMore} match: \texttt{[h, [e, [l, [l, [o] ] ] ] ]}.

Rewriting \texttt{OneOrMore} clauses into right-recursive form changes the structure of the parse tree into right-associative form. This transformation can be easily and automatically reversed once parsing is complete, by flattening each right-recursive linked list of \texttt{OneOrMore} subclause matches into an array of subclause matches of a single \texttt{OneOrMore} node. This is implemented in the reference parser by adding a \texttt{Match.getSubClauseMatches} method that flattens the right-recursive list for \texttt{OneOrMore} matches, simply returning the subclause matches as an array.

\subsection{\label{sec:lex}Reducing the number of spurious matches using a lex preprocessing pass}

When processing input bottom-up, spurious matches may be found that would not be found in a standard recursive descent parse, because a bottom-up parser is unaware of the higher-level context surrounding a given parsing position. For example, a bottom-up parse might spuriously match words inside of a comment or quoted string as identifiers. These matches will not be connected to the final parse tree, so they do not affect the structure of the final parse tree, but they do take up unnecessary space in the memo table, and creating and memoizing these spurious matches also wastes work.

Fortunately, spurious matches rarely cascade far up the grammar hierarchy, due to lacking structure recognizable by higher levels of the grammar; therefore, the impact on memory consumption and parsing time of these spurious matches tends to be limited. Spurious matches will therefore typically incur only a moderately small constant factor of overhead to parsing performance, and these spurious matches do not change the big-Oh time or space complexity of the parser (Section~\ref{sec:Performance}).

The overhead of spurious matches can be ameliorated through the use of a lexing or tokenization step, which greedily consumes tokens from left to right until all the input is consumed. Lexing is used as a preprocessing step by many parsers. A lexical token could be an identifier or symbol, or could be a longer match of a structured sequence, such as a complete quoted string or comment. These tokenized matches are then used to seed the memo table by the bottom-up parser, rather than seeding based on matches of all terminals at all input positions.

Note, however, that lexing will only work for simple grammars, and may cause recognition problems, since the limited pattern recognition capabilities of a lexer may change the recognized language. In fact it is impossible in the general case to create a lexer that is aware of all hierarchical context -- that is what a parser accomplishes. To allow a lexer to recognize all possible reasonable tokens, all matches of any lex token pattern at any position must be found. If multiple tokens of different length match at the same start position, then the lexer must look for the next token starting from each of the possible token end positions. In the worst case, this can devolve into having to match each lex token at each start position.

For these reasons, lexing is probably not advantageous or at least may not be advisable, unless the grammar is simple and lexing can be executed unambiguously.

Nevertheless, a ``lightweight lex'' preprocessing pass may still be advisable for performance reasons, for example to remove comments and to mark the input position ranges for all quoted strings before parsing, since these two cases in particular cause a performance impact for pika parsing.

\section{Automatic conversion of parse tree into an Abstract Syntax Tree (AST)}

The structure of the parse tree resulting from a parse is directly induced by the structure of the grammar, i.e. there is one node in the parse tree for each clause and subclause of the grammar that matched the input. Many of these nodes can be ignored (e.g. nodes that match semantically-irrelevant whitespace or comments in the input), and could be suppressed in the output tree. Mapping the parse tree into a simpler structure that contains only the nodes of interest results in the Abstract Syntax Tree (AST).

\begin{figure}[ht]
    \centering
    \includegraphics[width=1\textwidth]{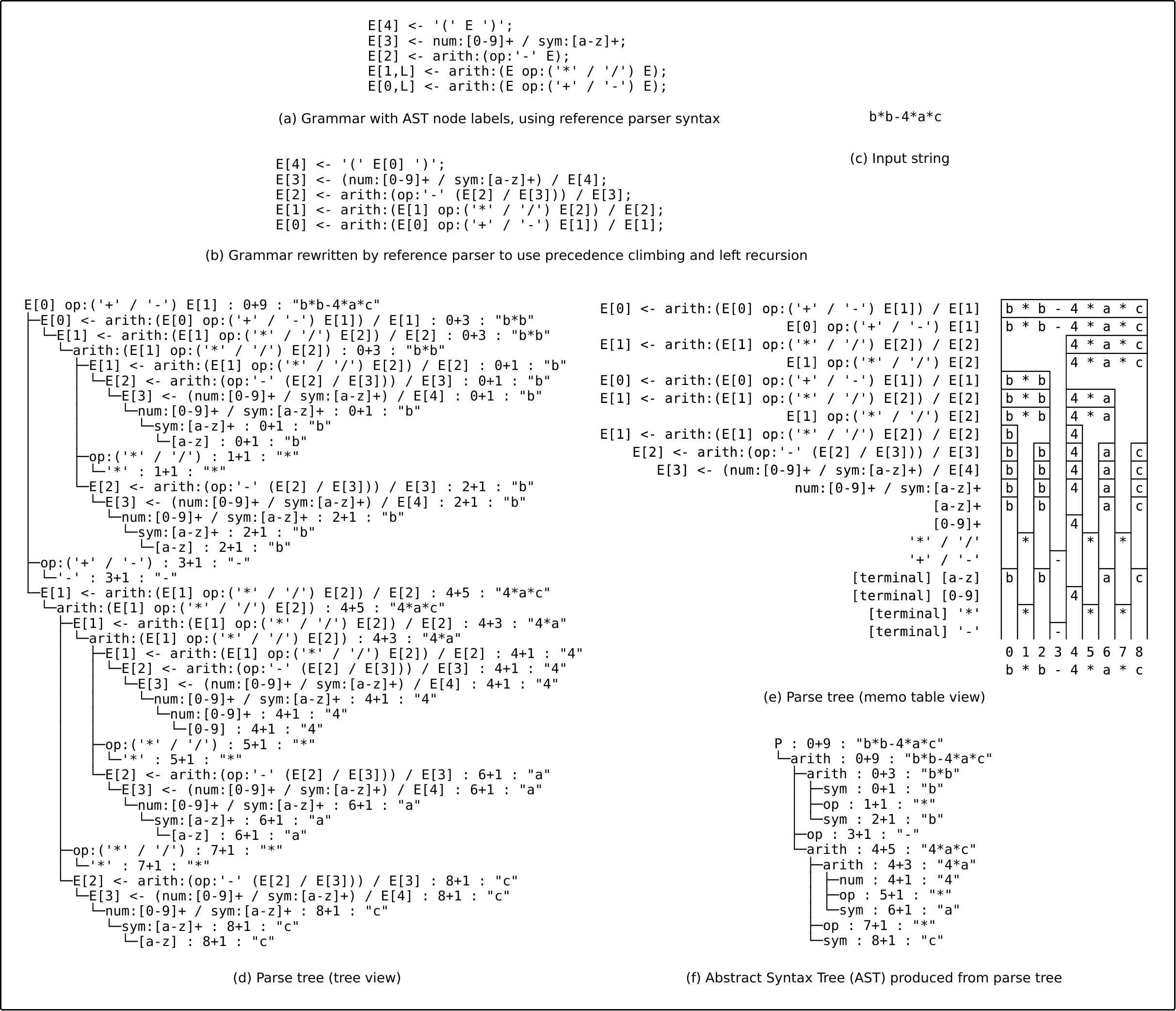}
    \caption{\label{fig:ast}Example of parse tree and abstract syntax tree produced by the reference pika parser}
    \Description[Example of parse tree and abstract syntax tree produced by the reference pika parser]{Example of parse tree and abstract syntax tree produced by the reference pika parser}
\end{figure}

The reference parser supports the syntax \texttt{(ASTNodeLabel:Clause)} for labeling any subclause in the grammar. After parsing is complete, all matches in the parse tree that do not have an AST node label are simply elided from the parse tree to create the AST (Fig.~\ref{fig:ast}).

\section{Future work}

The pika parser algorithm could be extended to enable \emph{incremental parsing}~\citep{dubroy2017}, in which when a document is parsed and then subsequently edited, only the minimum necessary subset of parsing work is performed, and as much work as possible is reused from the prior parsing operation.

To enable incremental parsing in a pika parser when a span of the input string that has changed, it is necessary to remove all matches from the memo table that span the changed region of the input. Bottom-up parsing is then seeded from terminal matches within the changed region, and the main parsing loop is run again in the usual bottom-up, right to left order.

Some care will also need to be taken to ensure that the memo table is indexed not based on start position of a match (which is an absolute position, therefore affected by insertions and deletions), but rather based on cumulative length of matches. Otherwise, insertions or deletions in the middle of the input string will require the start positions of all parse tree nodes to the right of the change to be updated with the new start position, and memo table entries will all have to be moved too. Dealing with this efficiently requires changing \texttt{MemoKey} so that it no longer includes an absolute \texttt{startPos} field, and instead memo entries would be indexed by column, from a linked list of columns corresponding to each input character.

Because the memo table tends to accumulate spurious matches that are not linked into the final parse tree, to keep memory usage low in an incremental pika parser that is built into an editor or IDE, the memo table should be periodically garbage collected by running a complete (non-incremental) parse from scratch.

\section{\label{sec:software}Reference implementation}

An MIT-licensed reference implementation of the pika parser is available at: \url{http://github.com/lukehutch/pikaparser}

The reference parser contains several optimizations not shown here, for example all clauses are \emph{interned}, so that rules that share a subclause do not result in duplicate rows in the memo table. Additionally, rule references are replaced with direct references to the rule, to save on lookup time while traversing the grammar during parsing. The optimizations for reducing the number of memoized zero-length matches described in Section~\ref{sec:zerolen} are also implemented in the reference parser. These optimizations make preprocessing the grammar more complicated, but result in significant performance gains.

The reference implementation includes a \emph{meta-grammar} or runtime parser generator that is able to parse a PEG grammar written in ASCII notation, making it easy to write new grammars.

\section{Conclusion}

The \emph{pika parser} is a new type of PEG parser that employs dynamic programming to parse a document in reverse (from the end to the beginning of the input), and bottom-up (from individual characters up to the root of the parse tree). This parsing order supports parsing of directly left-recursive grammars, making grammar writing simpler, and also enables almost perfect error recovery after syntax errors, making pika parsers useful for implementing IDEs and compilers. Pika parsers take time linearly proportional to the length of the input, and are very efficient for smaller grammars or long inputs. For large grammars, bottom-up parsing can incur a significant constant overhead per character of input, which may make pika parsers inappropriate for some performance-sensitive uses. Mechanisms for implementing precedence climbing and left or right associativity were demonstrated for PEG grammars, and several new insights were provided into precedence, associativity, and left recursion. A reference implementation of the pika parser is available under an MIT license.

\bibliographystyle{ACM-Reference-Format}
\bibliography{main}


\begin{thebibliography}{44}


\ifx \showCODEN    \undefined \def \showCODEN     #1{\unskip}     \fi
\ifx \showDOI      \undefined \def \showDOI       #1{#1}\fi
\ifx \showISBNx    \undefined \def \showISBNx     #1{\unskip}     \fi
\ifx \showISBNxiii \undefined \def \showISBNxiii  #1{\unskip}     \fi
\ifx \showISSN     \undefined \def \showISSN      #1{\unskip}     \fi
\ifx \showLCCN     \undefined \def \showLCCN      #1{\unskip}     \fi
\ifx \shownote     \undefined \def \shownote      #1{#1}          \fi
\ifx \showarticletitle \undefined \def \showarticletitle #1{#1}   \fi
\ifx \showURL      \undefined \def \showURL       {\relax}        \fi
\providecommand\bibfield[2]{#2}
\providecommand\bibinfo[2]{#2}
\providecommand\natexlab[1]{#1}
\providecommand\showeprint[2][]{arXiv:#2}

\bibitem[\protect\citeauthoryear{Aho, Johnson, and Ullman}{Aho
  et~al\mbox{.}}{1975}]%
        {aho1975}
\bibfield{author}{\bibinfo{person}{Alfred~V Aho}, \bibinfo{person}{Stephen~C
  Johnson}, {and} \bibinfo{person}{Jeffrey~D Ullman}.}
  \bibinfo{year}{1975}\natexlab{}.
\newblock \showarticletitle{Deterministic parsing of ambiguous grammars}.
\newblock \bibinfo{journal}{\emph{Commun. ACM}} \bibinfo{volume}{18},
  \bibinfo{number}{8} (\bibinfo{year}{1975}), \bibinfo{pages}{441--452}.
\newblock


\bibitem[\protect\citeauthoryear{Aho, Lam, Sethi, and Ullman}{Aho
  et~al\mbox{.}}{2006}]%
        {aho2006}
\bibfield{author}{\bibinfo{person}{Alfred~V. Aho}, \bibinfo{person}{Monica~S.
  Lam}, \bibinfo{person}{Ravi Sethi}, {and} \bibinfo{person}{Jeffrey~D.
  Ullman}.} \bibinfo{year}{2006}\natexlab{}.
\newblock \bibinfo{title}{Compilers: principles, techniques, and tools}.
\newblock
\newblock


\bibitem[\protect\citeauthoryear{Aho and Ullman}{Aho and Ullman}{1972}]%
        {aho1972}
\bibfield{author}{\bibinfo{person}{Alfred~V Aho} {and}
  \bibinfo{person}{Jeffrey~D Ullman}.} \bibinfo{year}{1972}\natexlab{}.
\newblock \bibinfo{booktitle}{\emph{The theory of parsing, translation, and
  compiling}}. Vol.~\bibinfo{volume}{1}.
\newblock \bibinfo{publisher}{Prentice-Hall Englewood Cliffs, NJ}.
\newblock


\bibitem[\protect\citeauthoryear{Aho and Ullman}{Aho and Ullman}{1977}]%
        {aho1977}
\bibfield{author}{\bibinfo{person}{Alfred~V. Aho} {and}
  \bibinfo{person}{Jeffrey~D. Ullman}.} \bibinfo{year}{1977}\natexlab{}.
\newblock \bibinfo{booktitle}{\emph{Principles of Compiler Design}}.
\newblock \bibinfo{publisher}{Addison-Wesley}.
\newblock


\bibitem[\protect\citeauthoryear{Chomsky}{Chomsky}{1956}]%
        {chomsky1956}
\bibfield{author}{\bibinfo{person}{Noam Chomsky}.}
  \bibinfo{year}{1956}\natexlab{}.
\newblock \showarticletitle{Three models for the description of language}.
\newblock \bibinfo{journal}{\emph{IRE Transactions on information theory}}
  \bibinfo{volume}{2}, \bibinfo{number}{3} (\bibinfo{year}{1956}),
  \bibinfo{pages}{113--124}.
\newblock


\bibitem[\protect\citeauthoryear{Chomsky}{Chomsky}{1959}]%
        {chomsky1959}
\bibfield{author}{\bibinfo{person}{Noam Chomsky}.}
  \bibinfo{year}{1959}\natexlab{}.
\newblock \showarticletitle{On certain formal properties of grammars}.
\newblock \bibinfo{journal}{\emph{Information and Control}}
  \bibinfo{volume}{2}, \bibinfo{number}{2} (\bibinfo{year}{1959}),
  \bibinfo{pages}{137--167}.
\newblock


\bibitem[\protect\citeauthoryear{Cocke, Schwartz, and
  of~Mathematical~Sciences}{Cocke et~al\mbox{.}}{1970}]%
        {cocke1970}
\bibfield{author}{\bibinfo{person}{William~John Cocke},
  \bibinfo{person}{Jacob~T Schwartz}, {and} \bibinfo{person}{Courant~Institute
  of Mathematical~Sciences}.} \bibinfo{year}{1970}\natexlab{}.
\newblock \bibinfo{booktitle}{\emph{Programming languages and their
  compilers}}.
\newblock \bibinfo{publisher}{Courant Institute of Mathematical Sciences.}
\newblock


\bibitem[\protect\citeauthoryear{Crockford}{Crockford}{2007}]%
        {crockford2007}
\bibfield{author}{\bibinfo{person}{Douglas Crockford}.}
  \bibinfo{year}{2007}\natexlab{}.
\newblock \showarticletitle{Top down operator precedence}.
\newblock \bibinfo{journal}{\emph{Beautiful Code: Leading Programmers Explain
  How They Think}} (\bibinfo{year}{2007}), \bibinfo{pages}{129--145}.
\newblock


\bibitem[\protect\citeauthoryear{Daniel}{Daniel}{1967}]%
        {younger1967}
\bibfield{author}{\bibinfo{person}{H Daniel}.} \bibinfo{year}{1967}\natexlab{}.
\newblock \showarticletitle{Younger. Recognition and parsing of context-free
  languages in time $n^3$}.
\newblock \bibinfo{journal}{\emph{Information and Control}}
  \bibinfo{volume}{10}, \bibinfo{number}{2} (\bibinfo{year}{1967}),
  \bibinfo{pages}{189--208}.
\newblock


\bibitem[\protect\citeauthoryear{de~Medeiros, Junior, and
  Mascarenhas}{de~Medeiros et~al\mbox{.}}{2020}]%
        {medeiros2020}
\bibfield{author}{\bibinfo{person}{S{\'e}rgio~Queiroz de Medeiros},
  \bibinfo{person}{Gilney de Azevedo~Alvez Junior}, {and}
  \bibinfo{person}{Fabio Mascarenhas}.} \bibinfo{year}{2020}\natexlab{}.
\newblock \showarticletitle{Automatic syntax error reporting and recovery in
  parsing expression grammars}.
\newblock \bibinfo{journal}{\emph{Science of Computer Programming}}
  \bibinfo{volume}{187} (\bibinfo{year}{2020}), \bibinfo{pages}{102373}.
\newblock


\bibitem[\protect\citeauthoryear{de~Medeiros and Mascarenhas}{de~Medeiros and
  Mascarenhas}{2018}]%
        {medeiros2018}
\bibfield{author}{\bibinfo{person}{S{\'e}rgio~Queiroz de Medeiros} {and}
  \bibinfo{person}{Fabio Mascarenhas}.} \bibinfo{year}{2018}\natexlab{}.
\newblock \showarticletitle{Syntax error recovery in parsing expression
  grammars}. In \bibinfo{booktitle}{\emph{Proceedings of the 33rd Annual ACM
  Symposium on Applied Computing}}. \bibinfo{pages}{1195--1202}.
\newblock


\bibitem[\protect\citeauthoryear{DeRemer}{DeRemer}{1969}]%
        {deremer1969}
\bibfield{author}{\bibinfo{person}{Franklin~Lewis DeRemer}.}
  \bibinfo{year}{1969}\natexlab{}.
\newblock \emph{\bibinfo{title}{Practical translators for LR (k) languages.}}
\newblock \bibinfo{thesistype}{Ph.D. Dissertation}.
  \bibinfo{school}{Massachusetts Institute of Technology}.
\newblock


\bibitem[\protect\citeauthoryear{Dubroy and Warth}{Dubroy and Warth}{2017}]%
        {dubroy2017}
\bibfield{author}{\bibinfo{person}{Patrick Dubroy} {and}
  \bibinfo{person}{Alessandro Warth}.} \bibinfo{year}{2017}\natexlab{}.
\newblock \showarticletitle{Incremental packrat parsing}. In
  \bibinfo{booktitle}{\emph{Proceedings of the 10th ACM SIGPLAN International
  Conference on Software Language Engineering}}. \bibinfo{pages}{14--25}.
\newblock


\bibitem[\protect\citeauthoryear{Earley}{Earley}{1970}]%
        {earley1970}
\bibfield{author}{\bibinfo{person}{Jay Earley}.}
  \bibinfo{year}{1970}\natexlab{}.
\newblock \showarticletitle{An efficient context-free parsing algorithm}.
\newblock \bibinfo{journal}{\emph{Commun. ACM}} \bibinfo{volume}{13},
  \bibinfo{number}{2} (\bibinfo{year}{1970}), \bibinfo{pages}{94--102}.
\newblock


\bibitem[\protect\citeauthoryear{Floyd}{Floyd}{1963}]%
        {floyd1963}
\bibfield{author}{\bibinfo{person}{Robert~W Floyd}.}
  \bibinfo{year}{1963}\natexlab{}.
\newblock \showarticletitle{Syntactic analysis and operator precedence}.
\newblock \bibinfo{journal}{\emph{Journal of the ACM (JACM)}}
  \bibinfo{volume}{10}, \bibinfo{number}{3} (\bibinfo{year}{1963}),
  \bibinfo{pages}{316--333}.
\newblock


\bibitem[\protect\citeauthoryear{Ford}{Ford}{2002a}]%
        {ford2002}
\bibfield{author}{\bibinfo{person}{Bryan Ford}.}
  \bibinfo{year}{2002}\natexlab{a}.
\newblock \emph{\bibinfo{title}{Packrat parsing: a practical linear-time
  algorithm with backtracking}}.
\newblock \bibinfo{thesistype}{Ph.D. Dissertation}.
  \bibinfo{school}{Massachusetts Institute of Technology}.
\newblock


\bibitem[\protect\citeauthoryear{Ford}{Ford}{2002b}]%
        {ford2002b}
\bibfield{author}{\bibinfo{person}{Bryan Ford}.}
  \bibinfo{year}{2002}\natexlab{b}.
\newblock \showarticletitle{Packrat parsing: simple, powerful, lazy, linear
  time, functional pearl}.
\newblock \bibinfo{journal}{\emph{ACM SIGPLAN Notices}} \bibinfo{volume}{37},
  \bibinfo{number}{9} (\bibinfo{year}{2002}), \bibinfo{pages}{36--47}.
\newblock


\bibitem[\protect\citeauthoryear{Ford}{Ford}{2004}]%
        {ford2004}
\bibfield{author}{\bibinfo{person}{Bryan Ford}.}
  \bibinfo{year}{2004}\natexlab{}.
\newblock \showarticletitle{Parsing expression grammars: a recognition-based
  syntactic foundation}. In \bibinfo{booktitle}{\emph{Proceedings of the 31st
  ACM SIGPLAN-SIGACT symposium on Principles of programming languages}}.
  \bibinfo{pages}{111--122}.
\newblock


\bibitem[\protect\citeauthoryear{Frost and Hafiz}{Frost and Hafiz}{2006}]%
        {frost2006}
\bibfield{author}{\bibinfo{person}{Richard~A Frost} {and}
  \bibinfo{person}{Rahmatullah Hafiz}.} \bibinfo{year}{2006}\natexlab{}.
\newblock \showarticletitle{A new top-down parsing algorithm to accommodate
  ambiguity and left recursion in polynomial time}.
\newblock \bibinfo{journal}{\emph{ACM SIGPLAN Notices}} \bibinfo{volume}{41},
  \bibinfo{number}{5} (\bibinfo{year}{2006}), \bibinfo{pages}{46--54}.
\newblock


\bibitem[\protect\citeauthoryear{Frost, Hafiz, and Callaghan}{Frost
  et~al\mbox{.}}{2007}]%
        {frost2007}
\bibfield{author}{\bibinfo{person}{Richard~A. Frost},
  \bibinfo{person}{Rahmatullah Hafiz}, {and} \bibinfo{person}{Paul~C.
  Callaghan}.} \bibinfo{year}{2007}\natexlab{}.
\newblock \showarticletitle{Modular and Efficient Top-down Parsing for
  Ambiguous Left-Recursive Grammars}. In \bibinfo{booktitle}{\emph{Proceedings
  of the 10th International Conference on Parsing Technologies}} (Prague, Czech
  REpublic) \emph{(\bibinfo{series}{IWPT ’07})}.
  \bibinfo{publisher}{Association for Computational Linguistics},
  \bibinfo{address}{USA}, \bibinfo{pages}{109–120}.
\newblock


\bibitem[\protect\citeauthoryear{Gray}{Gray}{1969}]%
        {gray1969}
\bibfield{author}{\bibinfo{person}{Jim Gray}.} \bibinfo{year}{1969}\natexlab{}.
\newblock \emph{\bibinfo{title}{Precedence parsers for programming languages}}.
\newblock \bibinfo{thesistype}{Ph.D. Dissertation}. \bibinfo{school}{University
  of California Berkeley, Calif.}
\newblock


\bibitem[\protect\citeauthoryear{Kasami}{Kasami}{1966}]%
        {kasami1966}
\bibfield{author}{\bibinfo{person}{Tadao Kasami}.}
  \bibinfo{year}{1966}\natexlab{}.
\newblock \showarticletitle{An efficient recognition and syntax-analysis
  algorithm for context-free languages}.
\newblock \bibinfo{journal}{\emph{Coordinated Science Laboratory Report no.
  R-257}} (\bibinfo{year}{1966}).
\newblock


\bibitem[\protect\citeauthoryear{Kay}{Kay}{1980}]%
        {kay1980}
\bibfield{author}{\bibinfo{person}{Martin Kay}.}
  \bibinfo{year}{1980}\natexlab{}.
\newblock \showarticletitle{Algorithm schemata and data structures in syntactic
  processing}.
\newblock \bibinfo{journal}{\emph{Technical Report}} (\bibinfo{year}{1980}).
\newblock


\bibitem[\protect\citeauthoryear{Kleene}{Kleene}{1951}]%
        {kleene1951}
\bibfield{author}{\bibinfo{person}{Stephen~Cole Kleene}.}
  \bibinfo{year}{1951}\natexlab{}.
\newblock \showarticletitle{Representation of events in nerve nets and finite
  automata}.
\newblock \bibinfo{journal}{\emph{RAND Research Memorandum}}
  \bibinfo{volume}{RM-704} (\bibinfo{year}{1951}).
\newblock


\bibitem[\protect\citeauthoryear{Knuth}{Knuth}{1962}]%
        {knuth1962}
\bibfield{author}{\bibinfo{person}{Donald~E Knuth}.}
  \bibinfo{year}{1962}\natexlab{}.
\newblock \showarticletitle{History of writing compilers}. In
  \bibinfo{booktitle}{\emph{Proceedings of the 1962 ACM National Conference on
  Digest of Technical Papers}}. \bibinfo{pages}{43}.
\newblock


\bibitem[\protect\citeauthoryear{Knuth}{Knuth}{1965}]%
        {knuth1965}
\bibfield{author}{\bibinfo{person}{Donald~E Knuth}.}
  \bibinfo{year}{1965}\natexlab{}.
\newblock \showarticletitle{On the translation of languages from left to
  right}.
\newblock \bibinfo{journal}{\emph{Information and Control}}
  \bibinfo{volume}{8}, \bibinfo{number}{6} (\bibinfo{year}{1965}),
  \bibinfo{pages}{607--639}.
\newblock


\bibitem[\protect\citeauthoryear{Kruskal}{Kruskal}{1983}]%
        {kruskal1983}
\bibfield{author}{\bibinfo{person}{Joseph~B Kruskal}.}
  \bibinfo{year}{1983}\natexlab{}.
\newblock \showarticletitle{An overview of sequence comparison: Time warps,
  string edits, and macromolecules}.
\newblock \bibinfo{journal}{\emph{SIAM review}} \bibinfo{volume}{25},
  \bibinfo{number}{2} (\bibinfo{year}{1983}), \bibinfo{pages}{201--237}.
\newblock


\bibitem[\protect\citeauthoryear{Lang}{Lang}{1974}]%
        {lang1974}
\bibfield{author}{\bibinfo{person}{Bernard Lang}.}
  \bibinfo{year}{1974}\natexlab{}.
\newblock \showarticletitle{Deterministic techniques for efficient
  non-deterministic parsers}. In \bibinfo{booktitle}{\emph{International
  Colloquium on Automata, Languages, and Programming}}. Springer,
  \bibinfo{pages}{255--269}.
\newblock


\bibitem[\protect\citeauthoryear{Levenshtein}{Levenshtein}{1966}]%
        {levenshtein1966}
\bibfield{author}{\bibinfo{person}{Vladimir~I Levenshtein}.}
  \bibinfo{year}{1966}\natexlab{}.
\newblock \showarticletitle{Binary codes capable of correcting deletions,
  insertions, and reversals}. In \bibinfo{booktitle}{\emph{Soviet Physics
  Doklady}}, Vol.~\bibinfo{volume}{10}. \bibinfo{pages}{707--710}.
\newblock


\bibitem[\protect\citeauthoryear{McNaughton and Yamada}{McNaughton and
  Yamada}{1960}]%
        {mcnaughton1960}
\bibfield{author}{\bibinfo{person}{Robert McNaughton} {and}
  \bibinfo{person}{Hisao Yamada}.} \bibinfo{year}{1960}\natexlab{}.
\newblock \showarticletitle{Regular expressions and state graphs for automata}.
\newblock \bibinfo{journal}{\emph{IRE transactions on Electronic Computers}}
  \bibinfo{number}{1} (\bibinfo{year}{1960}), \bibinfo{pages}{39--47}.
\newblock


\bibitem[\protect\citeauthoryear{Medeiros, Mascarenhas, and
  Ierusalimschy}{Medeiros et~al\mbox{.}}{2014}]%
        {medeiros2014}
\bibfield{author}{\bibinfo{person}{S{\'e}rgio Medeiros}, \bibinfo{person}{Fabio
  Mascarenhas}, {and} \bibinfo{person}{Roberto Ierusalimschy}.}
  \bibinfo{year}{2014}\natexlab{}.
\newblock \showarticletitle{Left recursion in parsing expression grammars}.
\newblock \bibinfo{journal}{\emph{Science of Computer Programming}}
  \bibinfo{volume}{96} (\bibinfo{year}{2014}), \bibinfo{pages}{177--190}.
\newblock


\bibitem[\protect\citeauthoryear{Myltsev}{Myltsev}{2013}]%
        {parboiled2}
\bibfield{author}{\bibinfo{person}{Alexander Myltsev}.}
  \bibinfo{year}{2013}\natexlab{}.
\newblock \bibinfo{booktitle}{\emph{Parboiled2: A Macro-Based PEG Parser
  Generator for Scala 2.12+}}.
\newblock
\urldef\tempurl%
\url{http://parboiled2.org}
\showURL{%
\tempurl}


\bibitem[\protect\citeauthoryear{Norvig}{Norvig}{1991}]%
        {norvig1991}
\bibfield{author}{\bibinfo{person}{Peter Norvig}.}
  \bibinfo{year}{1991}\natexlab{}.
\newblock \showarticletitle{Techniques for automatic memoization with
  applications to context-free parsing}.
\newblock \bibinfo{journal}{\emph{Computational Linguistics}}
  \bibinfo{volume}{17}, \bibinfo{number}{1} (\bibinfo{year}{1991}),
  \bibinfo{pages}{91--98}.
\newblock


\bibitem[\protect\citeauthoryear{Parr}{Parr}{2013}]%
        {antlr4}
\bibfield{author}{\bibinfo{person}{Terence Parr}.}
  \bibinfo{year}{2013}\natexlab{}.
\newblock \bibinfo{booktitle}{\emph{The definitive ANTLR 4 reference}}.
\newblock \bibinfo{publisher}{Pragmatic Bookshelf}.
\newblock
\urldef\tempurl%
\url{https://www.antlr.org}
\showURL{%
\tempurl}


\bibitem[\protect\citeauthoryear{Pingali and Bilardi}{Pingali and
  Bilardi}{2012}]%
        {pingali2012}
\bibfield{author}{\bibinfo{person}{Keshav Pingali} {and}
  \bibinfo{person}{Gianfranco Bilardi}.} \bibinfo{year}{2012}\natexlab{}.
\newblock \showarticletitle{Parsing with Pictures}.
\newblock \bibinfo{journal}{\emph{UTCS tech report TR-2012}}
  (\bibinfo{year}{2012}).
\newblock


\bibitem[\protect\citeauthoryear{Pratt}{Pratt}{1973}]%
        {pratt1973}
\bibfield{author}{\bibinfo{person}{Vaughan~R Pratt}.}
  \bibinfo{year}{1973}\natexlab{}.
\newblock \showarticletitle{Top down operator precedence}. In
  \bibinfo{booktitle}{\emph{Proceedings of the 1st Annual ACM SIGACT-SIGPLAN
  Symposium on Principles of Programming Languages}}. \bibinfo{pages}{41--51}.
\newblock


\bibitem[\protect\citeauthoryear{Ru{\v{z}}i{\v{c}}ka}{Ru{\v{z}}i{\v{c}}ka}{1979}]%
        {ruvzivcka1979}
\bibfield{author}{\bibinfo{person}{Peter Ru{\v{z}}i{\v{c}}ka}.}
  \bibinfo{year}{1979}\natexlab{}.
\newblock \showarticletitle{Validity test for Floyd's operator-precedence
  parsing algorithms}. In \bibinfo{booktitle}{\emph{International Symposium on
  Mathematical Foundations of Computer Science}}. Springer,
  \bibinfo{pages}{415--424}.
\newblock


\bibitem[\protect\citeauthoryear{Samelson and Bauer}{Samelson and
  Bauer}{1960}]%
        {samelson1960}
\bibfield{author}{\bibinfo{person}{Klaus Samelson} {and}
  \bibinfo{person}{Friedrich~L Bauer}.} \bibinfo{year}{1960}\natexlab{}.
\newblock \showarticletitle{Sequential formula translation}.
\newblock \bibinfo{journal}{\emph{Commun. ACM}} \bibinfo{volume}{3},
  \bibinfo{number}{2} (\bibinfo{year}{1960}), \bibinfo{pages}{76--83}.
\newblock


\bibitem[\protect\citeauthoryear{Sipser}{Sipser}{2012}]%
        {sipser2012}
\bibfield{author}{\bibinfo{person}{Michael Sipser}.}
  \bibinfo{year}{2012}\natexlab{}.
\newblock \bibinfo{booktitle}{\emph{Introduction to the Theory of
  Computation}}.
\newblock \bibinfo{publisher}{Cengage Learning}.
\newblock


\bibitem[\protect\citeauthoryear{Thomas}{Thomas}{1986}]%
        {thomas1986}
\bibfield{author}{\bibinfo{person}{J~Penello Thomas}.}
  \bibinfo{year}{1986}\natexlab{}.
\newblock \showarticletitle{Very fast LR parsing}.
\newblock \bibinfo{journal}{\emph{ACM SIGPLAN Notices}} \bibinfo{volume}{21},
  \bibinfo{number}{7} (\bibinfo{year}{1986}), \bibinfo{pages}{145--151}.
\newblock


\bibitem[\protect\citeauthoryear{Tomita}{Tomita}{2013}]%
        {tomita2013}
\bibfield{author}{\bibinfo{person}{Masaru Tomita}.}
  \bibinfo{year}{2013}\natexlab{}.
\newblock \bibinfo{booktitle}{\emph{Efficient parsing for natural language: a
  fast algorithm for practical systems}}. Vol.~\bibinfo{volume}{8}.
\newblock \bibinfo{publisher}{Springer Science \& Business Media}.
\newblock


\bibitem[\protect\citeauthoryear{Viterbi}{Viterbi}{1967}]%
        {viterbi1967}
\bibfield{author}{\bibinfo{person}{Andrew Viterbi}.}
  \bibinfo{year}{1967}\natexlab{}.
\newblock \showarticletitle{Error bounds for convolutional codes and an
  asymptotically optimum decoding algorithm}.
\newblock \bibinfo{journal}{\emph{IEEE transactions on Information Theory}}
  \bibinfo{volume}{13}, \bibinfo{number}{2} (\bibinfo{year}{1967}),
  \bibinfo{pages}{260--269}.
\newblock


\bibitem[\protect\citeauthoryear{Voisin}{Voisin}{1988}]%
        {voisin1988}
\bibfield{author}{\bibinfo{person}{Fr{\'e}d{\'e}ric Voisin}.}
  \bibinfo{year}{1988}\natexlab{}.
\newblock \showarticletitle{A bottom-up adaptation of Earley's parsing
  algorithm}. In \bibinfo{booktitle}{\emph{International Workshop on
  Programming Language Implementation and Logic Programming}}. Springer,
  \bibinfo{pages}{146--160}.
\newblock


\bibitem[\protect\citeauthoryear{Warth, Douglass, and Millstein}{Warth
  et~al\mbox{.}}{2008}]%
        {warth2008}
\bibfield{author}{\bibinfo{person}{Alessandro Warth}, \bibinfo{person}{James~R
  Douglass}, {and} \bibinfo{person}{Todd Millstein}.}
  \bibinfo{year}{2008}\natexlab{}.
\newblock \showarticletitle{Packrat parsers can support left recursion}. In
  \bibinfo{booktitle}{\emph{Proceedings of the 2008 ACM SIGPLAN symposium on
  Partial evaluation and semantics-based program manipulation}}.
  \bibinfo{pages}{103--110}.
\newblock


\end{thebibliography}

\newpage\section*{Listings}

\begin{lstlisting}[caption={Algorithm for topologically sorting clauses into bottom-up order},label={lst:topoSort}]
static void findReachableClauses(Clause clause, HashSet<Clause> visited, List<Clause> revTopoOrderOut) {
    if (visited.add(clause)) {
        for (var labeledSubClause : clause.labeledSubClauses) {
            var subClause = labeledSubClause.clause;
            findReachableClauses(subClause, visited, revTopoOrderOut);
        }
        revTopoOrderOut.add(clause);
    }
}

static void findCycleHeadClauses(Clause clause, Set<Clause> discovered, Set<Clause> finished,
        Set<Clause> cycleHeadClausesOut) {
    discovered.add(clause);
    for (var labeledSubClause : clause.labeledSubClauses) {
        var subClause = labeledSubClause.clause;
        if (discovered.contains(subClause)) {
            // Reached a cycle
            cycleHeadClausesOut.add(subClause);
        } else if (!finished.contains(subClause)) {
            findCycleHeadClauses(subClause, discovered, finished, cycleHeadClausesOut);
        }
    }
    discovered.remove(clause);
    finished.add(clause);
}

static List<Clause> findClauseTopoSortOrder(List<Rule> allRules, List<Clause> lowestPrecedenceClauses) {
    // Find toplevel clauses (clauses that are not a subclause of any other clause)
    var allClausesUnordered = new ArrayList<Clause>();
    var topLevelVisited = new HashSet<Clause>();
    for (var rule : allRules) {
        findReachableClauses(rule.labeledClause.clause, topLevelVisited, allClausesUnordered);
    }
    var topLevelClauses = new HashSet<>(allClausesUnordered);
    for (var clause : allClausesUnordered) {
        for (var labeledSubClause : clause.labeledSubClauses) {
            topLevelClauses.remove(labeledSubClause.clause);
        }
    }
    var dfsRoots = new ArrayList<>(topLevelClauses);
    // After toplevel clauses, start DFS from all lowest-precedence clause in each precedence hierarchy
    dfsRoots.addAll(lowestPrecedenceClauses);
    // Add toplevel clauses the set of all "head clauses" of cycles (all clauses reachable twice)
    var cycleDiscovered = new HashSet<Clause>();
    var cycleFinished = new HashSet<Clause>();
    var cycleHeadClauses = new HashSet<Clause>();
    for (var clause : topLevelClauses) {
        findCycleHeadClauses(clause, cycleDiscovered, cycleFinished, cycleHeadClauses);
    }
    for (var rule : allRules) {
        findCycleHeadClauses(rule.labeledClause.clause, cycleDiscovered, cycleFinished, cycleHeadClauses);
    }
    dfsRoots.addAll(cycleHeadClauses);
    // Topologically sort all clauses into bottom-up order
    var allClauses = new ArrayList<Clause>();
    var reachableVisited = new HashSet<Clause>();
    for (var topLevelClause : dfsRoots) {
        findReachableClauses(topLevelClause, reachableVisited, allClauses);
    }
    // Give each clause an index in the topological sort order, bottom-up
    for (int i = 0; i < allClauses.size(); i++) {
        allClauses.get(i).clauseIdx = i;
    }
    return allClauses;
}
\end{lstlisting}

\newpage

\begin{lstlisting}[caption={A primitive precedence-climbing grammar},label={lst:grammarIncomplete}]
E4 <- '(' E0 ')';
E3 <- ([0-9]+ / [a-z]+) / E4;
E2 <- ('-' E3) / E3;
E1 <- (E2 ('*' / '/') E2) / E2;
E0 <- (E1 ('+' / '-') E1) / E1;
\end{lstlisting}

\begin{lstlisting}[caption={Improved precedence-climbing grammar, supporting left associativity and self-nesting of parentheses and unary minus},label={lst:grammarFixed}]
E4 <- '(' E0 ')';
E3 <- ([0-9]+ / [a-z]+) / E4;
E2 <- ('-' (E2 / E3)) / E3;
E1 <- (E1 ('*' / '/') E2) / E2;
E0 <- (E0 ('+' / '-') E1) / E1;
\end{lstlisting}

\begin{lstlisting}[caption={Shorthand reference parser notation for the grammar of Listing~\ref{lst:grammarFixed}},label={lst:grammarRefParserSyntax}]
E[4] <- '(' E ')';
E[3] <- [0-9]+ / [a-z]+;
E[2] <- '-' E;
E[1,L] <- E ('*' / '/') E;
E[0,L] <- E ('+' / '-') E;
\end{lstlisting}

\begin{lstlisting}[caption={The \texttt{MemoKey} class},label={lst:MemoKey}]
class MemoKey implements Comparable<MemoKey> {
    Clause clause;
    int startPos;
}
\end{lstlisting}

\begin{lstlisting}[caption={The \texttt{Grammar} class (which contains the main \texttt{parse} method)},label={lst:Grammar}]
class Grammar {
    List<Clause> allClauses;

    MemoTable parse(String input) {
        var priorityQueue = new PriorityQueue<Clause>((c1, c2) -> c1.clauseIdx - c2.clauseIdx);
        var memoTable = new MemoTable(this, input);
        var terminals = new ArrayList<Clause>();
        for (var clause : allClauses) {
            if (clause instanceof Terminal && !(clause instanceof Nothing)) {
                terminals.add(clause);
            }
        }
        // Main parsing loop
        for (int startPos = input.length() - 1; startPos >= 0; --startPos) {
            priorityQueue.addAll(terminals);
            while (!priorityQueue.isEmpty()) {
                var clause = priorityQueue.remove();
                var memoKey = new MemoKey(clause, startPos);
                var match = clause.match(memoTable, memoKey, input);
                memoTable.addMatch(memoKey, match, priorityQueue);
            }
        }
        return memoTable;
    }
}
\end{lstlisting}

\begin{lstlisting}[caption={The \texttt{Clause} class},label={lst:Clause}]
abstract class Clause {
    Clause[] subClauses;
    List<Clause> seedParentClauses;
    boolean canMatchZeroChars;
    int clauseIdx;

    abstract void determineWhetherCanMatchZeroChars();
    abstract void addAsSeedParentClause();
    abstract Match match(MemoTable memoTable, MemoKey memoKey, String input);
}
\end{lstlisting}

\newpage

\begin{lstlisting}[caption={The implementation of the \texttt{Char} terminal (\texttt{Terminal} extends \texttt{Clause})},label={lst:Char}]
class Char extends Terminal {
    char c;    // The character to match

    @Override
    Match match(MemoTable memoTable, MemoKey memoKey, String input) {
        return memoKey.startPos < input.length() && input.charAt(memoKey.startPos) == c
            ? new Match(memoKey, /* len = */ 1, /* firstMatchingSubClauseIdx = */ 0,
                    /* subclauseMatches = */ new Match[0]);
            : null;
    }
}
\end{lstlisting}

\begin{lstlisting}[caption={The implementation of the \texttt{Seq} PEG operator},label={lst:Seq}]
class Seq extends Clause {
    @Override
    void determineWhetherCanMatchZeroChars() {
        // For Seq, all subclauses must be able to match zero characters for the whole clause to
        // be able to match zero characters
        this.canMatchZeroChars = true;
        for (var subClause : this.subClauses) {
            if (!subClause.canMatchZeroChars) {
                this.canMatchZeroChars = false;
                break;
            }
        }
    }
    
    @Override
    void addAsSeedParentClause() {
        // All sub-clauses up to and including the first clause that matches one or more characters
        // needs to seed its parent clause if there is a subclause match
        var added = new HashSet<>();
        for (var subClause : this.subClauses) {
            // Don't duplicate seed parent clauses in the subclause, if the subclause is repeated
            if (added.add(subClause)) {
                subClause.seedParentClauses.add(this);
            }
            if (!subClause.canMatchZeroChars) {
                // Subsequent subclauses don't need to seed this parent clause, because this match
                // consumes one character or more
                break;
            }
        }
    }

    @Override
    Match match(MemoTable memoTable, MemoKey memoKey, String input) {
        Match[] subClauseMatches = null;
        var currStartPos = memoKey.startPos;
        for (int subClauseIdx = 0; subClauseIdx < subClauses.length; subClauseIdx++) {
            var subClause = subClauses[subClauseIdx];
            var subClauseMemoKey = new MemoKey(subClause, currStartPos);
            var subClauseMatch = memoTable.lookUpBestMatch(subClauseMemoKey);
            if (subClauseMatch == null) {
                // Fail after first subclause fails to match
                return null;
            }
            if (subClauseMatches == null) {
                subClauseMatches = new Match[subClauses.length];
            }
            subClauseMatches[subClauseIdx] = subClauseMatch;
            currStartPos += subClauseMatch.len;
        }
        return new Match(memoKey, /* len = */ currStartPos - memoKey.startPos,
                /* firstMatchingSubClauseIdx = */ 0, subClauseMatches);
    }
}
\end{lstlisting}

\newpage

\begin{lstlisting}[caption={The \texttt{Match} class},label={lst:Match}]
class Match {
    MemoKey memoKey;
    int len;
    int firstMatchingSubClauseIdx;
    Match[] subClauseMatches;

    boolean isBetterThan(Match other) {
        // An earlier subclause match in a First clause is better than a later subclause match;
        // a longer match is better than a shorter match
        return (memoKey.clause instanceof First
                && this.firstMatchingSubClauseIdx < other.firstMatchingSubClauseIdx)
                        || this.len > other.len;
    }
}
\end{lstlisting}

\begin{lstlisting}[caption={The \texttt{MemoTable} class},label={lst:MemoTable}]
class MemoTable {
    Map<MemoKey, Match> memoTable = new HashMap<>();
    Grammar grammar;
    String input;

    Match lookUpBestMatch(MemoKey memoKey) {
        var bestMatch = memoTable.get(memoKey);
        if (bestMatch != null) {
            return bestMatch;
        } else if (memoKey.clause instanceof NotFollowedBy) {
            return memoKey.clause.match(this, memoKey, input);
        } else if (memoKey.clause.canMatchZeroChars) {
            return new Match(memoKey);
        }
        return null;
    }

    void addMatch(MemoKey memoKey, Match newMatch, PriorityQueue<Clause> priorityQueue) {
        var matchUpdated = false;
        if (newMatch != null) {
            var oldMatch = memoTable.get(memoKey);
            if ((oldMatch == null || newMatch.isBetterThan(oldMatch))) {
                memoTable.put(memoKey, newMatch);
                matchUpdated = true;
            }
        }
        for (var seedParentClause : memoKey.clause.seedParentClauses) {
            if (matchUpdated || seedParentClause.canMatchZeroChars) {
                priorityQueue.add(seedParentClause);
            }
        }
    }
}
\end{lstlisting}

\end{document}